# Understanding the Paradox of Primary Health Care Use: Empirical Evidence from India*

First Version: March 2021
This Version: July 2022

Pramod Kumar Sur
*Asian Growth Research Institute (AGI) and Osaka University*
pramodsur@gmail.com

Abstract

I study households' primary health care usage in India, which presents a *paradox*. I examine why most households use fee-charging private health care services even though (1) most providers have *no* formal medical qualifications and (2) in markets where qualified doctors offer *free care* through public hospitals. I present evidence that this puzzling practice has deep historical routes. I examine India's coercive forced sterilization policy implemented between 1976 and 1977. Utilizing the unexpected timing of the policy, multiple measures of forced sterilization, including at a *granular level*, and an instrumental variable approach, I document that places heavily affected by the policy have lower public health care usage today. I also show that the instrument I use is unrelated to a battery of demographic, economic, or political aspects before the forced sterilization period. Finally, I explore the mechanism and document that supply-side factors do not explain these differences. Instead, I demonstrate that places with greater exposure to forced sterilization have higher confidence in private hospitals and doctors to provide *good* treatment.

*I thank Abhijit Banerjee for teaching MITx course (MicroMasters Program in Data, Economics, and Development Policy) and clearly explaining the puzzles associated with primary health care usage in India, which become the primary motivation for this research. I also thank numerous seminar and conference participants for their excellent feedback. All remaining errors are my own. I am grateful for financial support from the Japan Society for the Promotion of Science (No. 21K13307). This paper is revised extensively and supersedes an earlier (and incomplete) version circulated under the title "Understanding the Puzzle of Primary Health Care Use: Evidence from India"

## 1. Introduction

The past decades have seen substantial efforts from national governments and international donor organizations to strengthen the public health care sector in developing countries. This is primarily due to the efforts to achieve the Millennium Development Goals (MDGs) and universal health coverage under Sustainable Development Goals (SDGs). However, despite these persistent efforts, the private sector still plays an important role in providing health care services. Moreover, a significant fraction of households in these countries continue to visit fee-charging private health care facilities for primary health care.[1] For instance, in 27 out of 32 lower-income countries, more than half of health care spending in 2018 was financed from private sources, and nearly 60% of the spending came from out-of-pocket expenditure (see Figure A1 in the Appendix).

In India, households' use of primary health care services presents a *paradox*.[2] The paradox is as follows. First, a significant fraction of households use fee-charging private health care services (CPR 2011; IIPS 2017; Peters et al. 2002).[3] Second, most private health care providers have no formal medical qualifications (Rohde and Viswanathan 1995; A. Banerjee, Deaton, and Duflo 2004; Das et al. 2022).[4] Although public health care providers are more qualified and offer (almost) free services, they have only around 20% of the market share (Muralidharan et al., n.d.). Third, while *lack of access* to public health care could partly explain the high use of the private sector, this cannot be the only explanation (see Figure 1). Strikingly, the private share of health care use is higher even in markets where qualified doctors offer free care in public hospitals; despite this service, the majority of health care visits are made to providers with no formal medical qualifications (Das, Holla, et al. 2016). Fourth, within India, there is a considerable variation in the types of health care usage across states (Peters et al. 2002; Muralidharan et al., n.d.). *Why does such a paradoxical situation exist?*

Figure 1 about here

[1] See for example, Bennett et al. (2005), Grépin (2016), and Mackintosh et al. (2016)
[2] Henceforth, I refer to "primary health care" as the "health care" for simplicity.
[3] For example, India has one of the highest proportions of private health spending anywhere in the world, constituting 82% of all health expenditure. Only five countries (Cambodia, the Democratic Republic of the Congo, Georgia, Myanmar, and Sierra Leone) have a higher dependence on private health financing (Peters et al. 2002).
[4] For example, according to the Indian Medical Association, about 1 million unqualified doctors (popularly known as quacks) practice allopathic medicine in India. https://www.ima-india.org/ima/free-way-page.php?pid=143#:~:text=It%20is%20estimated%20that%20about,and%20children%20are%20at%20stake. Accessed on January 28, 2021.

In this study, I contribute to this debate by providing a plausible reason for the existence of such a paradox. I question whether the current practice of health care use in India has deep historical routes. In particular, I study whether a domestic policy implemented by the government in the past explains this paradox. Combining contemporary household-level data with administrative archival records, I examine the aggressive family planning program implemented during the emergency rule in the 1970s and explore whether the coercion, disinformation, and carelessness involved in implementing the program could partly explain the paradox.

Between June 1975 and March 1977, India experienced a brief period of authoritarian rule.[5] This period, popularly known as "the emergency," which was proclaimed by then Prime Minister Indira Gandhi under the Indian constitution, suspended a wide range of civil liberties. As historians have argued and I will explain in detail in the next section, a unique policy that affected a majority of the population during this period was the introduction of an aggressive family planning program through forced sterilization (Panandiker, Bishnoi, and Sharma 1978; Gwatkin 1979; Dhar 2000; Nayar 2013; Williams 2014; Chandra 2017).[6] The policy, implemented in April 1976, led to a sharp increase in the number of sterilizations (see Figure A2 in the Appendix). More than 8 million sterilizations were performed in a single year between April 1976 and March 1977, more than three times the previous year's figure. Historical records, court rulings, and anecdotal evidence suggest that these sterilization targets were accomplished through incentives and disincentives, coercion, disinformation, carelessness, and fear (Shah Commission of Inquiry 1978; Panandiker, Bishnoi, and Sharma 1978).

My main hypothesis is that the forced sterilization policy may have had unintended effects on future health care usage in India. There are reasons to believe that the policy could have unintended consequences. First, all sterilizations, primarily administered through coercion and disincentives, were performed by government (public) doctors in public hospitals or temporary sterilization camps established by the government. Due to intense pressure, targets to meet, and carelessness, no aftercare was administered, which sometimes led to serious side effects, including death. For example, according to the report published by the Indian Government, 1,778 complaints of deaths related to sterilization were registered during this period (Shah Commission of Inquiry

[5] The authoritarian rule officially ended in March 1977. However, it was substantially relaxed in January 1977.

[6] Henceforth, I refer to "the aggressive family planning program through forced sterilization" as the "forced sterilization policy" or "the policy" for simplicity.

1978). Second, false information was delivered by public health care workers to motivate individuals to be sterilized. In a survey of four Indian states (Bihar, Madhya Pradesh, Punjab, and Uttar Pradesh) during the forced sterilization period, Panandiker, Bishnoi, and Sharma (1978) summarize the types of disinformation provided to motivate sterilization acceptors:

> *"What was often told was that sterilization, vasectomy or tubectomy, is a simple, quick and safe operation which stops child birth permanently.* […] *To the more circumspect of the prospective clients, it was also quietly added that in case of need for a child-birth later it could be* ***reversed also****. Nobody explained how an operation is performed, in what manner it stops the conception and what its consequences are to the health of a person.* […] *As the program was generally time and target bound, their mission was "Quick Catch" rather than to carry conviction." (p. 104)*

Considering these insights, I examine the consequences of the forced sterilization policy on India's future health care usage behavior.

My main finding is that the forced sterilization policy has had a direct impact on explaining the puzzling practice of health care use in India. To test my hypothesis, I use data from India's national representative National Family and Health Survey in 2015–16 (NFHS-4) that reports the sources of households' health care use. To measure the exposure to the forced sterilization policy, I use sterilization statistics from the historical yearbooks published by the Ministry of Health and Family Planning, Government of India. I document that higher exposure to the forced sterilization policy is associated with lower use of public health care facilities today.

I bolster this interpretation with several exercises. First, I show that my results remain *quantitatively* similar if I include a wide range of household, geographic, and health facility level controls in my regression. Second, I present evidence that the results are also robust to a number of alternative definitions of exposure to the forced sterilization policy, including the total number of sterilizations performed in 1976–77, the excess number of sterilizations performed in 1976–77, total and excess sterilizations on a natural logarithm scale, and an alternative measure of exposure to sterilization measured by male sterilization, which constituted about 75% of the sterilizations performed during this period.

Next, I turn to the task of addressing concerns of reverse causality and omitted variable bias using an instrumental variable (IV) estimation approach. To identify the causal impact, I need an instrument that exogenously determines the sterilization performance during this period. For this, I exploit the unique history of the implementation of the forced sterilization policy and use distance from New Delhi to state capitals as an instrument. This instrument was first developed and empirically tested in Sur (2021) to examine the impact of the forced sterilization on vaccination coverage across India.

The instrument is constructed considering a well know insight from the emergency period that the forced sterilization policy was aggressively undertaken owing to the active role played by Sanjay Gandhi, the younger son of Prime Minister Indira Gandhi (Gwatkin 1979; Indian National Congress 2011; Nayar 2013; Chandra 2017). As historians have argued, which I will explain in detail later, due to Mr. Gandhi's personal influence, sterilization was aggressively undertaken in the northern parts of India. As a result, distance from New Delhi, which was *previously irrelevant*, emerged as an important determinant of excess sterilizations and is itself capable of explaining two-thirds of the variation in sterilization performance among the states (Gwatkin 1979). Based on these insights, I use distance from New Delhi to state capitals as an instrument to capture the variation in exposure to the forced sterilization policy.

The IV regression produces estimates that are similar to the ordinary least squares (OLS) estimates. The estimated effects are quantitatively sizable and significant. For example, I find that an average increase in excess sterilizations (from zero to about 3.2 times) decreases the use of public health care facilities today by about 18.6 percentage points. This is relative to a sample mean of 44.2% for my sample as a whole. It suggests that an average increase in excess sterilizations can explain up to *a 42 percent decline* in public health care use in India.

There are several potential threats to my identification strategy. For one, distance from New Delhi to state capitals may be related to earlier sterilization performance across states. Or the effects of distance from New Delhi may be working through a different mechanism, for instance, because the demographic characteristics are positively associated with states closer to the capital. This is because the primary reason for implementing the coercive family planning program was to check the increasing population growth in India. I thus perform an extensive set of falsification exercises to empirically test the identifying assumption and bolster the validity of my approach and interpretation. First, I present evidence that my instrument is unrelated to a battery of earlier sterilization performances across states. Moreover, I do not find a consistent pattern of excess female sterilization performed in 1976-77, which was not the main focus during this period. Second, I show that the source of variation I exploit is unlikely to be confounded by other competing explanations. For example, my instrument is not related to population across states from the 1971 census (before the forced sterilization period), rural population in 1971, the share of the Muslim population in 1971, or population growth between 1961 and 1971. Third, I document that the states closer to New Delhi are not systematically different based on development characteristics (measured by net domestic per-capita, labor force participation rate, and the share of the population working in the organized sector across states), and thus bolstering my overall causal mechanism. Fourth, I present evidence that distance from the national capital (New Delhi) does not predict voting behavior towards PM Indira Gandhi's Indian National Congress (INC) party before the forced sterilization period, and provide additional evidence that political characteristics are not systematically associated with my instrument as well.

Finally, as an alternative and complementary strategy, I use two other sources of variation in forced sterilization available at a more *granular* level. Several scholars have argued that the forced sterilization became the biggest political issue in 1977 election in India and played an important role in the defeat of Indira Gandhi's INC Party (see for example, Banerjee and Duflo 2011; Gwatkin 1979; Jaffrelot and Anil 2021; Weiner 1978).[7] Therefore, I use the constituency level variation in vote share of INC party in 1977 election (in absolute term) and the change in its vote share between 1977 and 1971 (in relative term) as my alternative measure that directly captures the variation in forced sterilization performed just before the national election in 1977. In both cases the results are precise and consistent with my interpretation that forced sterilization is associated with lower usage of public health care facilities today.

I also explore the potential mechanisms. I first examine the reasons why households do not use public health care facilities. An *obvious* reason could be the supply-side factors. I test whether this is the case. I use the data from NFHS-4 that asks an additional question to households who do not use public health care facilities, asking them to report the reasons. I find that the effects of exposure to the forced sterilization policy on standard supply-side constraints—such as no nearby public health facility, inconvenient timing, absence of health personnel, and long waiting time—are minimal, sometimes negative, and statistically insignificant. This suggests that supply-side factors are less likely to be the reasons for higher usage of private health care facilities. However, higher exposure to the forced sterilization policy has a positive and statistically significant effect on households responding to "poor quality of care" and "other" as their reasons for not using public health care facilities.

[7] I will explain this in more detail in Section 5.

Next, I delve further into the reasons why households respond with "poor quality of care" and "other" as their reasons for not using public health care facilities. Recall that, during the forced sterilization period, public health care workers did not provide appropriate medical care and often delivered false information to motivate individuals to undergo sterilization. I, therefore, check whether loss of confidence in providing proper treatment is a direct plausible channel. I use data from the Indian Human Development Survey-II (a national representative household survey conducted between 2011 and 2012) to examine how exposure to forced sterilization policy affects confidence in hospitals and doctors. I document that households belonging to states highly exposed to the forced sterilization policy exhibit a *higher* level of confidence in private hospitals and doctors in providing good treatment. In contrast, they exhibit a *lower* level of confidence in government hospitals and doctors. These findings are somehow *puzzling* because, as I have noted earlier, most private health care providers in India have no formal medical qualifications to practice medicine. Overall, the results imply that a lower level of confidence of government hospitals and doctors in providing proper treatment—presumably due to historical reasons—is a plausible mechanism for explaining the puzzling practice of health care usage in India.

In addition to the historical literature discussed previously, this paper builds on and contributes to a diverse range of works in economics. First, I contribute to the rich and active literature on understanding the factors associated with health care usage in developing countries in general and in India, in particular. Several studies have documented that supply-side determinants are contributing factors for lower usage of public health care facilities (Peters et al. 2002; Banerjee, Deaton, and Duflo 2004; De Costa and Diwan 2007; Das, Holla, et al. 2016).[8] Furthermore, owing to the higher usage of private health care facilities, there have been recent debates about training unqualified private health care practitioners to achieve better health care delivery (Government of Telangana 2015; Das, Chowdhury et al. 2016). However, limited evidence exists on why people use private health care facilities in the first place, *especially* in markets with a qualified doctor offering *free* care in a public hospital. Additionally, we have limited systematic evidence about the pathways through which social or historical characteristics influence households' decisions to use health care services. I build on and contribute to this growing literature in three ways. First, I provide an empirical investigation of the importance of history in shaping current health care usage. Second, I offer plausible causal evidence that historical characteristics—i.e., domestic policies implemented in the past—indeed influence current decision-making for households' health care usage. Third, I provide *mechanism* as well as the *reasons for the mechanism* for this puzzling practice today.

[8] Supply side factors include lack of nearby health care facility, transport network, absenteeism, and quality of health care among others.

Health care provision is a public good, and universal health coverage is considered to be an integral part of the Sustainable Development Goals (Goal 3, Target 3.8). To achieve universal health care coverage in developing countries, international organizations such as the World Bank advocate delivering health care through free or nominally priced medical care in publicly run facilities staffed by qualified doctors (World Bank 2003). However, a significant fraction of households in these countries *still* visits fee-charging private health care providers (Bennett et al. 2005; Grépin 2016; Mackintosh et al. 2016). Furthermore, households in developing countries spend a substantial portion of their resources on health care (World Bank Group 2019). This paper builds on and contributes to the literature on understanding a *potential* reason for this practice. I present evidence suggesting that health intervention through government policies implemented in the past could have a long-term and persistent effect on explaining such health-seeking behavior at present. In particular, I present evidence that the erosion of confidence in public health care providers due to historical policies is an important reason for such puzzling practice. Nonetheless, my focus on the historical determinants of difference in health care usage should not imply that other factors are unimportant. A number of existing studies have shown the importance of determinants such as supply-side constraints, service quality, culture, and information asymmetry as important factors contributing to private health care usage in developing countries (see Dupas (2011), Das and Hammer (2014), Dupas and Miguel (2017), for a detailed review in this field). As I demonstrate here, a potential historical legacy affecting the usage of health care remains even today.

The remainder of the paper is structured as follows. Section 2 provides a brief background of the emergency and the forced sterilization implemented during this period. Section 3 explains the historical and contemporary data used in the empirical analysis. Section 4 presents the OLS, IV results, and falsification tests to show the validity of the instrument. Section 5 discusses the estimates using two alternative measures of variation in forced sterilization available at a more granular level. Section 6 discusses the mechanisms, and Section 7 concludes. The Appendix provides additional robustness checks and results.

## 2. Context: Emergency Rule and Forced Sterilizations in India

In this section, I provide a brief background of the emergency rule period and forced sterilization policy in India. For a detailed overview of the emergency period, please see Nayar (2013) and Dhar (2018). Furthermore, for a detailed overview of the sterilization program implemented during this period, please see Panandiker, Bishnoi, and Sharma (1978), Shah Commission of Inquiry (1978), and Gwatkin (1979).

On June 25, 1975, Prime Minister Indira Gandhi declared a national emergency under Article 352 of the Indian constitution.[9] The exact reason for the declaration of emergency rule is controversial to this day. However, sociologists, political scientists, and historians argue that a combination of economic and political difficulties concerning her leadership and India are the most credible factors.

The emergency rule allowed Ms. Gandhi to suspend a wide range of civil liberties under the Indian constitution. Thousands of people, including key opposition leaders, were arrested, the press was censored, and public gatherings and strikes were declared illegal. With all the power in Ms. Gandhi's hands, she undertook a series of constitutional amendments and introduced new legislation to govern the country. The executive power of the emergency allowed the central government to give directions to states as to the manner in which the executive power was to be exercised. However, on January 23, 1977, Ms. Gandhi unexpectedly called for an election in March of that year. She released the opposition leaders from jail, lifted press censorship, and permitted public meetings once again. The emergency period officially ended in March after the Indian National Congress Party (INC) was defeated in the Lok Sabha election (the lower house of the Indian parliament).

[9] Article 352 (1) states that "If the President is satisfied that a grave emergency exists whereby the security of India or of any part of the territory thereof is threatened, whether by war or external aggression or armed rebellion, he may, by Proclamation, make a declaration to that effect in respect of the whole of India or of such part of the territory thereof as may be specified in the Proclamation Explanation. A Proclamation of Emergency declaring that the security of India or any part of the territory thereof is threatened by war or by external aggression or by armed rebellion may be made before the actual occurrence of war or of any such aggression or rebellion, if the President is satisfied that there is imminent danger thereof."

A hallmark of the emergency period was an aggressive family planning program through sterilization. It was launched in April 1976, about ten months after the proclamation of the emergency. The aggressive family planning program started with the New Population Policy (NPP) introduced to the parliament by the union minister of Health.[10] The NPP mainly concentrated on propagating sterilization as its method of family planning. Temporary sterilization camps were established by the government. With the NPP's introduction, the central government authorized and endorsed a series of coercive measures for sterilization and, in extreme cases, the provision for compulsory sterilization. The central and state governments substantially increased the financial rewards for sterilization acceptors. Through a range of incentives and disincentives, they pressured their employees to get sterilized and to motivate others to do so. In certain cases, quotas were imposed at the district level. Additionally, state and central government employees were given quotas to produce people for sterilization. In other cases, citizens were required to produce sterilization certificates to access basic facilities, such as public health care, irrigation, and subsidized food through ration cards (Shah Commission of Inquiry 1978; Panandiker, Bishnoi, and Sharma 1978).

The aggressive nature of the family planning program and the concentration of effort on propagating sterilization resulted in about 8.3 million sterilizations between April 1976 and March 1977, more than three times the number in the previous year. During the peak, over 1.7 million sterilizations were performed in September 1976 alone, a figure that equaled the annual average for the ten preceding years (Gwatkin 1979). The majority of the sterilizations, about 75%, involved men undergoing vasectomies.

Historical records, court rulings, and previous studies suggest that incentives and disincentives were provided, sterilization quotas were imposed, and coercion was applied to motivate individuals to undergo sterilization during this period.[11] For example, in Uttar Pradesh, a motivation bonus—of 6 rupees (about 0.7 US dollars in 1976) per person motivated to undergo sterilization—was provided to the family planning health care staff for each person sterilized in excess of their quota. Additionally, as a form of disincentive, over 24,000 public health care employees were not paid their wages in June 1976 for their failure to complete their quotas for the

[10] For a detail overview of the NPP, see Singh (1976).
[11] For a detailed discussion on quota enforcement, incentives and disincentives, coercion, and fear around sterilization during the emergency, see Panandiker, Bishnoi, and Sharma (1978) and Shah Commission of Inquiry (1978).

April–June quarter (Panandiker, Bishnoi, and Sharma 1978). Some extreme rules were also made. In a letter from the Chief Secretary of Bihar (the most senior position in the civil services of the states in India), Divisional Commissioners were informed of the following decision:

> *Non-achievement of targets would render officers and staff of Health Department liable to punishment, e.g., censure in case of achievement short of a cent* (100) *per cent, stoppage of increment with cumulative effect if achievement was less than 75 per cent and termination of service if achievement fell short of 50 per cent.* (Shah Commission on Inquiry 1978 p. 172)

Anecdotal evidence suggests that individuals were influenced and disinformed to lead them to accept sterilization during this period. In a survey of four Indian states, Panandiker, Bishnoi, and Sharma (1978) found that about 72 percent of the sterilized people were motivated by the influence of government officials, and more than 58 percent were influenced by health care personnel.[12] Only about 19 percent underwent sterilization on their own initiative, and the remaining 9 percent were motivated by friends and relatives. ***None*** of those surveyed underwent sterilization because of the lure of money, and no one cited any case where the money had played a motivating part. They noted the following environment in which most individuals underwent sterilization:

> *The common sites for the* (sterilization) *camps in the rural areas were big villages, locations where village festivals and fairs were held, including weekly markets, and sometimes the primary health centers themselves. In the towns, the camps were generally held near the crowded localities inhabited by the lower middle and poor class people. Preparations for the camps were made well in advance. Mobile units of medical staff were deputed to perform the operations. Family planning field staff would go round the neighboring villages or localities, usually in government vehicles, to exhort and "persuade" people to come forward for sterilization. Revenue officials, block staff, and school teachers were also often pressed into service for mobilizing people for operation at the camps, and generally free transport—trucks, pick-ups, etc.—were provided to carry people to camp-sites. At the camps, the assembled people were given refreshments, usually tea and snacks, before operation, and care was taken that nobody* ***slipped away****. [...] Every acceptor, [...],*

[12] The four Indian states are Bihar, Madhya Pradesh, Punjab, and Uttar Pradesh.

*was also given a cash award at the time of his or her discharge from the camp. (Panandiker, Bishnoi, and Sharma 1978, pp. 108–111).*

The aggressive nature of the program led to serious consequences, including medical complications, death, and sterilization of ineligible individuals. Once a person was sterilized and allowed to go home, he or she was generally forgotten and left to fend for himself or herself if any complications arose. Due to increased pressure, targets to meet, and carelessness, no aftercare was administered, which sometimes led to serious side effects, including death. According to the report published by the Indian Government, 1,778 complaints of deaths related to sterilization were registered during this period. In several instances, ineligible individuals were sterilized as well. For example, reports of about 548 sterilizations of unmarried individuals had been registered during this period. Similarly, in Uttar Pradesh, 11,434 individuals with fewer than two children and 69 persons over 55 years were sterilized (Shah Commission of Inquiry 1978).

This was the first major program since India's independence in which the people were pitted against the government. Every action of the government under the sterilization program was regarded as suspect and created a credibility gap in the government's relationship with the people. The levels of coercion, disinformation, and carelessness associated with sterilization during this period gave free scope for the spread of rumors and fears. As a result, many people tried to avoid being caught by sterilization programs. Whenever a sterilization campaign was launched or a camp held, a warning was spread through word of mouth to distant places and among a large number of people *"Nasbandi-wale aarahe hein, Hoshiya rahena, Bhai"* (The sterilization operators are coming. **Beware brother**) (Panandiker, Bishnoi, and Sharma 1978).

The legacy of forced sterilization remained in peoples' minds and was evident even after the emergency rule ended. For example, the sterilization program became the biggest political issue and played an important role in the subsequent elections in March 1977 and the defeat of Indira Gandhi's Indian National Congress (INC) party (Banerjee and Duflo 2011; Gwatkin 1979; Weiner 1978; Williams 2014). Indeed, *nasbandi* (the term used for "sterilization" in India) became the focal point of the 1977 election campaign, and the INC's vote share declined substantially in places that had been deeply affected by the forced sterilization drives (Weiner 1978).

The new government formed in 1977 immediately reversed the forced sterilization policy. Additionally, to repair the poor reputation of the health ministry, the Indian Government changed the name from the 'Ministry of Health and Family Planning' to the 'Ministry of Health and Family

Welfare.' In the post-emergency period, the family planning program shifted from vasectomy to tubectomy, with women becoming the primary target (Basu 1985). The word "emergency" itself became synonymous with "sterilization," and, even today, individuals refer to the emergency period as the period of sterilization (Tarlo 2000). The emergency rule remains controversial today and is considered one of the darkest periods in the history of Indian democracy.

## 3. Data Sources and Description

My dataset constitutes historical administrative data, two recent nationally representative household survey data (NFHS-4 and IHDS-II), and other contemporary but more aggregated data on population and health care facilities.

### *3.1. Historical Data on Sterilization*

The historical data on sterilization for this paper comes from the historical yearbooks published by the Ministry of Health and Family Planning, Department of Family Planning, Government of India. The yearbooks report yearly statistics on family planning programs performed between April and March every year, along with various demographic and health statistics. Notably, the historical yearbooks include the number of sterilizations performed and the types of sterilization performed at the state level.

I digitized and use the sterilization data from the historical yearbooks published by the Ministry of Health and Family Planning. Figure A2 presents the total number of sterilizations along with the types of sterilization performed in India every year since the start of the program in 1956. As the figure shows, there is a sharp increase in the total number of sterilizations performed during 1976–77. It is also evident that most sterilizations performed during this period were vasectomies.

Figure 2 presents the total number of sterilizations performed between April 1976 and March 1977 at the state level. To provide a visual representation, I group the sterilization measures into several broad categories, with darker shades denoting a greater number of sterilizations performed. As we can see, there is a considerable variation in the exposure to the forced sterilization policy at the state level. As I will explain in detail in my IV analysis, a key determinant for this variation was the unique history of this period and the important role played by Sanjay Gandhi, the son of the prime minister.

Figure 2 about here

### *3.2. Contemporary Household Survey Data*

I combine the historical data on exposure to the forced sterilization policy with two nationally representative household survey datasets from India—the National Family and Health Survey in 2015–16 (NFHS-4) and the Indian Human Development Survey-II in 2011–12 (IHDS-II). The NFHS-4 is a stratified two-stage sample that covers all Indian states and union territories. The IHDS-II surveys cover all states and union territories of India, with the exception of the Andaman and Nicobar Island and Lakshadweep.

My primary outcome variable is the data on households' sources of health care from the NFHS-4. The NFHS-4 asks households about the source of health care that they generally use when household members become sick.[13] It categorizes health care sources into four broad groups: the public health sector, nongovernmental organizations (NGO) or trust hospitals/clinics, the private health sector, and others. I construct an indicator variable measuring whether the household members generally use the public health sector. In the NFHS-4 sample, about 45% of households report using public health care facilities.[14] In Figure 3, I present the percentage of households who generally use public health care facilities at the state level. As we can see, there is a wide variation in the use of public health care facilities at the state level. This is consistent with the findings of Peters et al. (2002) and Muralidharan et al. (n.d.), who found that there is a large variation in the types of health care used across states.

Figure 3 about here

I use additional data to examine the mechanism through which the forced sterilization policy influences decision-making concerning health care utilization. My first additional outcome variables to explore this mechanism are the responses in the NFHS-4 concerning reasons why households do not use public health care facilities. Respondents are allowed to provide multiple answers to this question in the survey. It reports a total of six reasons: no nearby facility, facility timing not convenient, health personnel often absent, waiting time too long, poor quality of care, and other reasons. I consider each possible reason separately as my outcome of interest to understand the factors that affect a household's intention to avoid using public health care facilities.

[13] The question the NFHS-4 asks is "When members of your household get sick, where do they generally go for treatment?".

[14] This number is weighted by sample weights. The unweighted figure is about 47%.

My second additional outcome variable to explore the reasons for the mechanism is the data on confidence in institutions from the Indian Human Development Survey-II in 2011–12 (IHDS-II). The IHDS-II asks households questions on their confidence in hospitals and doctors to provide good treatment. It asks separate questions concerning government hospitals and doctors and private hospitals and doctors. The respondents can choose between three possible answers: a great deal of confidence, only some confidence, and hardly any confidence at all. The IHDS-II assigns the value 1 to "a great deal of confidence," 2 to "only some confidence," and 3 to "hardly any confidence at all." Therefore, a higher score constitutes a lower level of confidence.

### *3.3 Other Data*

I constructed two other sources of variation in forced sterilization at a granular level—vote share of INC party in 1977 election (in absolute term) and the change in its vote share between 1977 and 1971 elections (in relative term). The data on the INC party's vote share is referenced from the election results of the Lok Sabha—the lower house of the Indian parliament. The data come from the statistical reports published by the Election Commission of India.[15] The data on general election results are available at the parliamentary constituency levels. I use household's geocoded location in NFHS-4 clusters and match them with the assembly constituency of India before delimitation of boundaries in 2008. I then construct parliamentary constituency based on the shape file of the assembly constituencies of India (see Figure A3 in the Appendix for the geographical distribution of NFHS-4 clusters matched with assembly constituencies).[16]

I additionally collect data on an extensive set of controls. I use aggregate data on population and health care facilities and personnel to control for potential covariates that could affect both the exposure to forced sterilization and current health care utilization. I collect population data from the 2011 population census to construct state-level population densities. Additionally, I collect health care facility and health care personnel data from Rural Health Statistics to construct information on hospitals and on doctors per 1,000 people at the state level.

Finally, I collect historical data on demographic and development indicators to examine the validity of my instrument. I collect state-level demographic data on total population, rural population, the share of the Muslim population, and the population growth rate from the historical census. Additionally, I collect historical data on domestic per-capita, labor force participation rate, and the share of workers in the organized sector as proxy measures for development across states.

---

[15] See https://eci.gov.in/statistical-report/statistical-reports/

[16] The assembly constituency level shape file is provided by Sandip Sukhtankar (See https://uva.theopenscholar.com/sandip-sukhtankar/data). The shape file contains the name of the parliamentary constituency for which each assembly constituency belong to. I acknowledge Manasa Patnam and Sandip Sukhtankar for their generous effort and providing the shape file for free.

## 4. Main Results

### *4.1. Descriptive Evidence and OLS Estimation*

I begin by showing a simple relationship between historical exposure to the forced sterilization policy and India's current health care use through a scatter plot. Figure 4 presents the association between the total number of sterilizations performed in 1976–77 (expressed in 100,000s of individuals) and the percentage of households who generally use public health care facilities calculated from the NFHS-4 at the state level. In Figure A4 in the Appendix, I present the same correlation plot but scale the symbols so that the sizes represent the population of the state (from the 2011 census) for better visualization. As we can see, the sterilization performance in 1976–77 is strongly associated with less use of public health care facilities. Additionally, it appears to be very general and not driven by a small number of influential outliers (such as population of a state like UP).

Figure 4 about here

Then, I examine this relationship by controlling for household, geographic, and health care characteristics that are potentially important determinants of a household's health care utilization. My baseline estimating equation is:

$$Y_{hcs} = \alpha + \beta Forced\ Sterilization_s + \gamma_1 X^H_{hcs} + \gamma_2 X^C_{cs} + \gamma_3 X^S_s + \epsilon_{hcs} \qquad (1),$$

where $h$ indexes households, $c$ denotes NFHS-4 clusters, and $s$ denotes states. The variable $Y_{hcs}$, denotes my outcome variable, which varies at the household level $h$. It is an indicator variable that measures whether the household usually uses public health care facility. The variable $Forced\ Sterilization_s$ denotes one of my measures of exposure to the forced sterilization policy in state *s*. I will discuss this variable in more detail below. $X^H_{hcs}$, $X^C_{cs}$, and $X^S_S$ are vectors of household-level, NFHS-4 cluster-level, and state-level control variables, respectively.

The household-level control variables $X^{H}_{hcs}$ include age and sex of the household head, household size, nine religion fixed effects, four caste fixed effects, 21 education of the household head fixed effects, four household wealth index fixed effects, an indicator for whether the household has a below poverty line (BPL) card, and an indicator for whether any household member is covered by health insurance. These controls are intended to proxy for household income and wealth. $X^{C}_{cs}$ is a vector of NFHS-4 cluster-level covariates intended to capture the characteristics of the place where the household lives, such as altitude in meters, altitude squared, and an indicator of whether the cluster is urban. $X^{S}_{S}$ is a vector of covariates meant to capture state-level characteristics that are likely to be correlated with the use of public health care facilities. They include population density per square kilometer (in log), hospitals per 1,000 people, and doctors per 1,000 people. $\epsilon_{hcs}$ is a random error term, capturing all omitted factors, which I allow to be heteroscedastic and correlated across households; in practice, the standard errors I report in my main analysis are clustered at the state level. Because NFHS-4 is a stratified two-stage sample designed to produce indicators at the district, state, and national levels and separate estimates for urban and rural areas, undersampling and oversampling are observed in many places. To account for this issue, I will conduct the regression analysis using weights defined in the NFHS-4.

I present the OLS estimates of equation (1) in Table 1. In column 1, I use the total number of sterilizations performed in a state in 1976–77 (expressed in 100,000 individuals) as my measure of the intensity of the forced sterilization policy. The estimated coefficient for $Forced\ Sterilization_s$, β, is negative and statistically significant. This is consistent with my hypothesis that the forced sterilization has a negative effect on households' usage of public health care facilities.

Table 1 about here

A possible concern with the above estimation is that the distribution of my explanatory variable—Total Sterilizations Performed in 1976–77 (in 100,000)—is right-skewed with a large number of observations taking on small values. We can observe this from Figure A5, which plots the histogram of the number of sterilizations performed in 1976–77 at the state level. To account for this issue, I estimate equation (1) using the natural log of the number of sterilizations performed in 1976–77 as my measure of the intensity of the forced sterilization policy. I present the estimates in column 2 of Table 1. The results are similar to column 1, as I find a significant negative correlation between this measure of forced sterilization and the usage of public health care facilities.

In columns 1 and 2, I use the total number of sterilizations performed in 1976–77 to measure exposure to the forced sterilization policy. One potential limitation of this measure is that it does not account for the number of sterilizations that would have happened anyway in the absence of the NPP under which the forced sterilization policy was undertaken. Accounting for this difference is important because sterilization, as a family planning method, has been performed in India since the 1950s, as shown in Figure A2. In column 3, I account for this issue and use an alternative measure of the forced sterilization policy measured by excess sterilizations performed in 1976–77 over and above the 1975–76 numbers.[17] Additionally, in column 4, I report estimates using the natural log of the excess number of sterilizations performed in 1976–77. As we see, the results are similar using these alternative sterilization measures.

Finally, I report the estimates considering a better measure of forced sterilization policy that collectively accounts for India's emergency rule, the size of states, and the state-level historical characteristics associated with sterilization performance. The estimates reported in columns 3 and 4 use the absolute number of sterilizations to measure forced sterilization policy. Some shortcomings of these measures are that they (1) do not account for the difference in the size of states and (2) do not account for any state-level historical factors associated with the level of sterilization performance that I do not capture in the estimation. To account for these issues, in column 5, I report the estimates after normalizing the excess sterilizations performed using sterilization figures in the previous year (1975–76). Specifically, I define $Forced\ Sterilization_s$ as follows:

$$Excess\ Sterilization_s = \frac{\#\ of\ sterilization\ in\ (1976{\sim}77)_s - \#\ of\ sterilization\ in\ (1975{\sim}76)_s}{\#\ of\ sterilization\ in\ (1975{\sim}76)_s}$$

I normalized the previous year's figures to account for the effect of emergency rule in India (as 1975–76 was part of the emergency period) and isolate the impact of the forced sterilization policy from India's emergency rule.[18] This is because the emergency rule itself could affect the outcome in several ways, given that India was primarily governed by autocratic rule during this period, and that it involved numerous policy changes. As we can see, the results remain robust to this alternative specification, as shown in column 5.

[17] Using alternative measures of excess sterilization performed in 1976–77, involving deducting the average of the last two years or three years, produces nearly identical results.
[18] Normalizing by the average of the last two years or three years as an alternative measure produces nearly identical results.

In Section B of the Appendix, I present a series of robustness and sensitivity checks. I briefly discuss them here. First, I verify whether my results are sensitive to the inclusion and exclusion of controls. To verify this, I report estimates that involve adding each set of controls sequentially for each of my measures of forced sterilization (Tables B1–B5). In addition, I check whether the results remain robust when considering excess vasectomies only as an alternative measure of $Forced\ Sterilization_s$ (Table B6), given that vasectomies constituted the majority of sterilization operations (see Figure A2). My findings are robust to these alternative specifications and different measures of the forced sterilization policy.

For the remainder of the analysis, I use state-level excess sterilizations performed in 1976–77 normalized by the 1975–76 sterilization figure as my baseline measure of exposure to the forced sterilization policy (the specification from column 5 of Table 1). This provides the best measure as it accounts for India's emergency rule and is normalized by both size and state-level historical characteristics associated with sterilization performance. However, as I illustrate in Table 1, my results are not reliant on this choice of measure only.

### *4.2. Instrumental Variable Analysis*

In section 4.1, I found that the forced sterilization policy has a negative association with the use of public health care facilities today. In this section, I address concerns of reverse causality and omitted variable bias using an IV approach. To identify a plausible causal impact, I need an instrument that exogenously determines the sterilization performance during this period. For this, I exploit the *unique* history of the implementation of the forced sterilization policy and use distance from New Delhi to state capitals as an instrument to capture the state-level variation in exposure to the excess sterilizations performed during the emergency rule in India.

The unique history of the implementation of the forced sterilization policy is as follows. As described by Gwatkin (1979), Nayar (2013), and Chandra (2017), among others, the aggressive manner in which the forced sterilization policy was conducted was due to the active role of Sanjay Gandhi, the younger son of the then Prime Minister Indira Gandhi. Although he had not been officially elected and held no official position, Sanjay Gandhi rapidly rose to power during the emergency period. Family planning was a key element of his self-declared five-point program that became the central theme of his public addresses.[19] Mr. Gandhi and some of his close colleagues in Delhi were at the center of the action and continuously influenced regional political leaders and bureaucrats, particularly those in the states adjacent to the national capital of Delhi (Shah Commission of Inquiry 1978). Owing to his personal influence, sterilization was aggressively undertaken in the northern part of India, particularly in the states adjacent to New Delhi. As a result, distance from New Delhi, which was previously *irrelevant*, emerged as an important determinant of performance in the excess sterilization program and is itself capable of explaining two-thirds of the variation in performance among the states (Gwatkin 1979). This unique history of the implementation of the forced sterilization policy during the emergency period and the personal influence of Sanjay Gandhi provide a basis for the construction and the validity of my instrument.

I report the IV estimates in Table 2, including each set of control variables sequentially across columns 1–4. Panel A reports the first-stage estimates for the instrument. The first-stage estimates show that distance from New Delhi to state capitals is negatively correlated with excess sterilizations performed during the emergency rule in India. This is consistent with the general narrative and Gwatkin's (1979) observation. In panel B, I present the second-stage estimates.[20] They suggest a negative and statistically significant effect of the forced sterilization policy on the current use of public health care facilities. In Section C of the Appendix, as a robustness check, I consider excess vasectomies only (Table C1). The estimates are robust to this alternative specification and similar to the results reported in Table 2.

Table 2 about here

Not only are the negative coefficient estimates of Table 2 statistically significant, but they are also economically meaningful. Column 4 of Table 2 indicates that an average increase in excess sterilizations—where sterilizations increased on average by about 3.2 times compared with the rates prior to the enforced sterilization policy—decreases the use of public health care facilities today by about 18.6 percentage points. This is relative to a sample mean of 44.2% for our sample as a whole. It suggests that the forced sterilization policy has a large effect—of about 42%—on the use of public health care facilities in India.

### *4.3. Threats against Instrument Validity*

[19] The other four programs were adult education, abolition of dowries, planting of trees, and eradication of the caste system.
[20] Additionally, I also report the reduced form estimates in Panel C.

There are several concerns about the validity of my instrument. In particular, the IV strategy rests on the assumption that the instrument I use—distance from New Delhi to the state capital—is exogenous and satisfies the exclusion restriction. I provide some qualitative evidence, including Gwatkin (1979), supporting that my instrument is driven by the personal influence of the son of the then prime minister and in particular, it is not correlated with sterilization performance previously. In this section, I perform an extensive set of falsification exercises to empirically test the identifying assumption and examine the validity of my approach and interpretation.

My first falsification exercise consists of examining sterilization performance before 1976. Because Sanjay Gandhi had no personal influence over sterilization before 1976, my IV—if exogenous—should have no predictive power on sterilization performance before 1976. First, in panel A of Figure 5, I illustrate Gwatkin's insight. In particular, I present the relationship between my instrument and excess sterilizations performed in 1975–76, the year immediately *before* the implementation of the forced sterilization policy. As we can see, the simple scatter plot suggests no association between distance from New Delhi to state capitals and excess sterilizations performed in 1975–76. I formally test this relationship by estimating several placebo exercises. Precisely, I examine whether distance from New Delhi to states capitals can explain the excess sterilization performed in the last four years. I also present the associations disaggregated by excess vasectomies (male sterilization) and tubectomies (female sterilization). I present the results in Figure 6. As we can see, distance from New Delhi to state capitals do not predict the excess sterilizations performed in the previous year's well. In particular, we find fairly precise close to zero estimates for these outcomes. Additionally, the comparison to the effects on the excess sterilization in 1976–77, shown at the top (in red), indicates that the quantitative magnitude of this impact is also extremely small.

Figure 5 about here

Figure 6 about here

I undertake a second falsification exercise to test whether the forceful nature of the sterilization *'only,'* is related to my instrument. To test the hypothesis, I focus my attention on excess female sterilizations, or tubectomies performed during this period, which were not the main focus of the forced sterilization program (Shah Commission of Inquiry 1978; Gwatkin 1979; Basu 1985). The forced sterilization program did not focus on female sterilization because tubectomies constitute major surgery and require longer hospitalization for recovery. Conversely, vasectomies are relatively quick to perform, and recipients can be discharged on the same day of the operation. During the emergency period, most sterilizations were performed in temporary camps. The existing infrastructure struggled to cope with the large number of operations induced by the increased pressure and targets imposed, which was another reason why tubectomy was not the main focus during this period.

This analogy provides a falsification test for my instrument. I visually present the relationship between my instrument and excess tubectomies performed in 1976–77 in panel B of Figure 5. As we can see, the relationship is flat. We find no association between distance from New Delhi to state capitals and excess tubectomies performed in 1976–77. I then test this relationship by formally estimating a placebo exercise in row 11 of Figure 6 (presented in blue color). The result suggests that my instrument does not have any predictive power for excess female sterilizations performed during the forced sterilization period.

As a third step, I use an extensive set of demographic and development indicators before and during the forced sterilization period and present evidence that my instrument is not systematically associated with these outcomes. I test for historical demographic indicators because the primary reason for implementing the coercive family planning program was to check the increasing population growth in India. Additionally, I test for development indicators to examine whether the difference in development characteristics across states is associated with my instrument. I present the evidence in the lower panel of Figure 6 (expressed in green color). As we can observe, all the demographic and development indicators are not associated with the instrument. In particular, I find that the state-level total population, rural population, the share of the Muslim population in 1971, and population growth rate between 1961 and 1971 are not systematically associated with my instrument. In addition, I also present evidence that more developed states (as expressed by higher net domestic per capita, labor force participation rate, and the share of the population working in the formal sector) are not associated with distance from New Delhi. These tests confirm that my instrument is also orthogonal to a large number of pre-forced sterilization era demographic and development characteristics.

A final concern is that distance from New Delhi to state capitals, even if orthogonal to pre-1976 state-level sterilization, demographic, and development characteristics may be working through other channels. The most important alternative that comes to mind here is that this instrument may be correlated with voting behavior in the 1971 election in which Indira Gandhi became the Prime Minister and later instituted the authoritarian rule in India. I investigate whether this is the case. I look at the vote share of the INC party received in the 1971 election and present the results in the last row of Figure 6 (expressed in purple color). As we can observe, there is no evidence of a statistical association between my instrument and the INC's vote share in the 1971 election.

Overall, I find no evidence of a higher level of sterilizations before 1976-77 or a greater level of female sterilizations during the forced sterilization period in states closer to New Delhi. Furthermore, I tested a battery of demographic, development, and political indicators before and during the forced sterilization period. I found no evidence of any systematic difference closer to the national capital. These extensive set of falsification exercises increases our confidence both in the validity of my instrument and, more importantly, in the specific channel via which this instrument is hypothesized to impact the excess sterilization during the latter part of emergency rule in India.

## 5. Results with Alternative Sources of Variation in Forced Sterilization at a Granular Level

My main hypothesis—that forced sterilization contributed to the lower usage of public health care facilities—would also suggest that alternative measures of variation explaining forced sterilization should have similar effects on health care usage in India. I now investigate this question looking at the effect of INC party's vote share in 1977 parliamentary election.

As noted earlier, several scholars, including Banerjee and Duflo (2011), Gwatkin (1979), Weiner (1978), and Williams (2014), have extensively argued that the draconian forced sterilization policy played an important role in Indira Gandhi and her INC party's defeat in the 1977 parliament election. Indeed, sterilization became the primary political issue of the 1977 election campaign. In particular, INC's vote share declined primarily in places that had been deeply affected by the forced sterilization.[21]

[21] See Figure A6 in the appendix for the association between excess sterilization and INC's vote share. Figure A6(A) shows that excess sterilization is negatively associated with INC's vote share in 1977. Comfortingly, FigureA6(B) documents that the excess sterilization is uncorrelated with INC's vote share in 1971—the immediate last major parliament election to 1977. For a detail econometric analysis of the legacy of forced sterilization on voting behavior in India, including the pre-trend, see Sur (2022).

Table 3 shows results exploiting this source of variation at a granular level.[22] The first four columns present the results considering the constituency level variation in INC's vote share in 1977 (absolute term). As we can see there is a precisely estimated positive impact of INC's vote share in 1977 (suggesting lower level of forced sterilization) on public health care use. The next four columns present the change in INC's vote share between 1977 and 1971 (relative term). The implied quantitative magnitudes are similar to those we saw in the first four columns. Finally, the results are similar and remain robust if I conduct an IV analysis considering distance from New Delhi as the instrument (see Table C2 in the Appendix).

Table 3 about here

Overall, all the results presented until now are consistent with my key hypothesis—those places heavily affected by the forced sterilization led to the decline in the usage of public health care facilities.

## 6. Investigating the Mechanism

In the previous sections, I found that the forced sterilization policy has had a negative and sizable effect on public health care use in India. In this section, I examine plausible channels or mechanisms that explain this negative effect. First, I explore the reasons provided by the households in the NFHS-4 questionnaire. Then, I examine confidence in health care facilities and doctors as a plausible direct mechanism.

### *6.1. Examining the Reasons Given in the NFHS-4*

The NFHS-4 asks households who do not use public health care facilities to explain the reasons why. It offers a total of six reasons: no nearby facility, facility timing not convenient, health personnel often absent, waiting time too long, poor quality of care, and other reasons. Respondents were allowed to select multiple answers. I consider each answer separately as outcomes of interest to understand whether the forced sterilization policy has had any effect on households selecting these answers as reasons for not visiting a public health care facility.[23]

[22] As we can see, the number of observations drops in Table 3. This is primarily because the INC party did not contest its candidates from all the parliamentary constituencies in 1971 and 1977 elections.
[23] Estimating the effects by indexing the reasons is difficult as respondents are allowed to choose multiple answers.

I present the results in Table 4. As shown, the effects of exposure to the forced sterilization policy on standard supply-side factors—such as no nearby facility, facility timing not convenient, health personnel often absent, and waiting time too long—are minimal, sometimes negative, and statistically insignificant. These estimates suggest that supply-side constraints are not the mechanism explaining why households do not use public health care facilities in areas where exposure to the sterilization policy was high.

Table 4 about here

However, column 5 suggests that higher exposure to the forced sterilization policy has a positive and significant effect on households selecting “poor quality of care” as their reason for not using public health care facilities. Additionally, the estimates in column 6 suggest that households are more likely to answer “other” as their reasons for not using public health care facilities in states where sterilization exposure was higher.

In Section D of the Appendix, I present a series of robustness and sensitivity checks. I first verify whether my results are sensitive to the inclusion and exclusion of controls (Table D1). Second, I check my results for robustness when considering excess vasectomies only (Table D2). As we can see, overall, the estimates are robust to these alternative specifications and similar to the results reported in Table 3.

### *6.2. Examining Confidence in Health Care Facilities and Doctors from IHDS-II Data*

In this section, I delve further into plausible reasons for households answering “poor quality of care” and “other” as their reasons for not using public health care facilities. I check whether loss of confidence in public health care and public health care personnel is a plausible channel for the current pattern of health care use. Several studies have shown that health interventions in the past are associated with subsequent mistrust in medicine (Alsan and Wanamaker 2018; Martinez-Bravo and Stegmann 2021; Lowes and Montero 2021).

I test for this channel in Indian context because, as I noted earlier, false information was delivered by public health care workers to motivate individuals to undergo sterilization during this period. Additionally, during the forced sterilization period, after being sterilized and discharged from the camp or hospital, patients were generally left to fend for themselves even if any complications arose, which led to serious side effects for some, including death. Therefore, I check whether loss of confidence is a plausible channel for the current avoidance of public health care.

I use data from the Indian Human Development Survey-II in 2011–12 (IHDS-II) on confidence in institutions to answer this question. The IHDS-II asks households separate questions on confidence in government hospitals and doctors and private hospitals and doctors to provide good treatment. The respondents can choose between three possible answers to which the IHDS-II assign values of 1, 2, and 3, respectively: "a great deal of confidence", "only some confidence", and "hardly any confidence at all".

Figure 7 presents the association through scatter plots to aid visual understanding. In panel (A), I plot the relationship between excess sterilizations in 1976–77 and confidence in government hospitals and doctors. In panel (B), I additionally plot the correlation between excess sterilizations in 1976–77 and confidence in private hospitals and doctors. We see a positive association in panel (A) and negative association in panel (B). It suggests that households belonging to states highly exposed to the forced sterilization policy exhibit a lower level of confidence in government hospitals and doctors, and a higher level of confidence in private hospitals and doctors in providing good treatment.

Figure 7 about here

Next, I examine this relationship through an IV regression in Table 5. In column 1, I estimate the relationship between the forced sterilization policy and confidence in government hospitals and doctors. Additionally, in column 2, I estimate the relationship between the forced sterilization policy and confidence in private hospitals and doctors. As we can see, the results are similar to the association we found in Figure 7. The results imply that a lower level of confidence in or distrust towards government hospitals and doctors is a plausible reason for lower usage of public health care facilities. In Section E of the Appendix, I report a series of alternative analyses showing that the results are robust overall and similar to those reported in Table 5.

Table 5 about here

## 7. Conclusion

In this paper, I study households' primary health care use in India, which presents a paradox. I examined the importance of a domestic policy, implemented by the government in the past, in shaping current health care use in India. In particular, I examined whether the aggressive family planning program under which a forced sterilization policy was implemented during the period of emergency rule in the 1970s could partly explain the lower use of public health care facilities today.

Using data from the NFHS-4, I examined households' source of health care. I found that greater exposure to the forced sterilization policy is associated with lower use of public health care facilities today. I also found that the results were robust to a variety of controls, a number of alternative measures of exposure to the forced sterilization policy including at a granular level, and when examining the impact through an IV approach.

Next, I examined the plausible mechanisms. First, I examined the reasons why households do not use public health care facilities. I found that the exposure to the forced sterilization policy did not have significant effects on standard supply-side constraints. However, higher exposure to the forced sterilization policy has had a large, positive, and significant effect on households answering that "poor quality of care" and "other" reasons were why they chose not to use public health care facilities.

I delved further into the reasons why households responded that "poor quality of care" or "other" reasons led them to avoid public health care facilities. Using data from the IHDS-II on confidence in institutions, I found that households belonging to states that were highly exposed to the forced sterilization policy exhibit a lower (higher) level of confidence in public (private) hospitals and doctors in providing good treatment. These results imply that a lower level of confidence in public hospitals and doctors is a plausible mechanism for lower usage of public health care facilities. This is expected given that public health care staff provided disinformation to motivate individuals to accept sterilization and did not provide proper aftercare during the sterilization period which led to serious complications including death.

My results provide robust evidence suggesting that historical policies implemented by the government in the past have had a strong and persistent impact on shaping health-seeking behavior today. This has important implications for understanding the puzzling factors behind the higher demand for private and unqualified health care services, even in markets where public provider exist and offer free health care through qualified doctors. I also offer mechanisms for this puzzling practice and provide plausible reasons for these mechanisms to prevail and persists in the long run. Overall, they highlight the unintended consequences associated with domestic health policy in the past and more importantly, the importance of understanding such contexts for the design and implementation of public policy and future interventions.

A key question is how we can generalize this historical event in India to other contexts. The most direct parallel comparison can be to the countries that have implemented such coercive domestic policies in the past. Coercive domestic policies, especially in the health care sector, are not uncommon. For instance, forced sterilization policy, as a direct comparison to the Indian experience, has also been implemented in several developing as well as developed countries, including Bangladesh, Brazil, China, Japan, Kenya, South Africa, United States, and Uzbekistan to name a few (see Reilly (2015) for a detailed review). Peru's forced sterilization program during Alberto Fujimori's regime and Uzbekistan governments' recent policy on sterilizing women are some excellent examples that can plausibly fit the Indian context directly. Additional research is needed to understand whether such coercive policies have untended consequences, particularly on health care usage, as national governments and international donor organizations are massively investing in the public health care sector to achieve universal health coverage under Sustainable Development Goals.

Figure 1: Use of Public and Private Health Care Facilities based on *Access* to Public Health Care

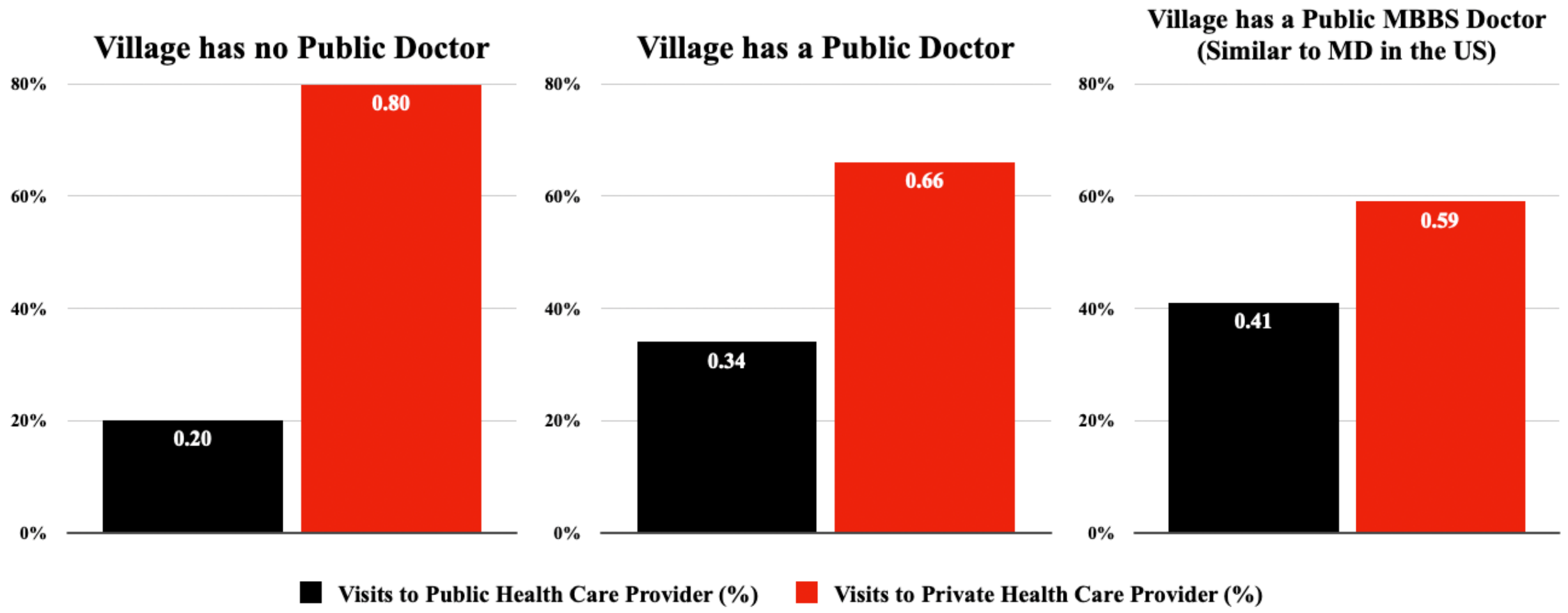

*Notes:* The figure presents the proportion of Indian households who visit public and private facilities for primary health care based on the availability of public health care facilities in a village. For example, the use of public health care increases from 20% to 34% in villages where there is a public doctor. In villages with a public MBBS doctor (similar to MD in the US), the usage rate is still low at only 41%.
Data Source: The MAQARI Project.

Figure 2: Number of Sterilizations Performed in 1976-77 (in 100,000s)

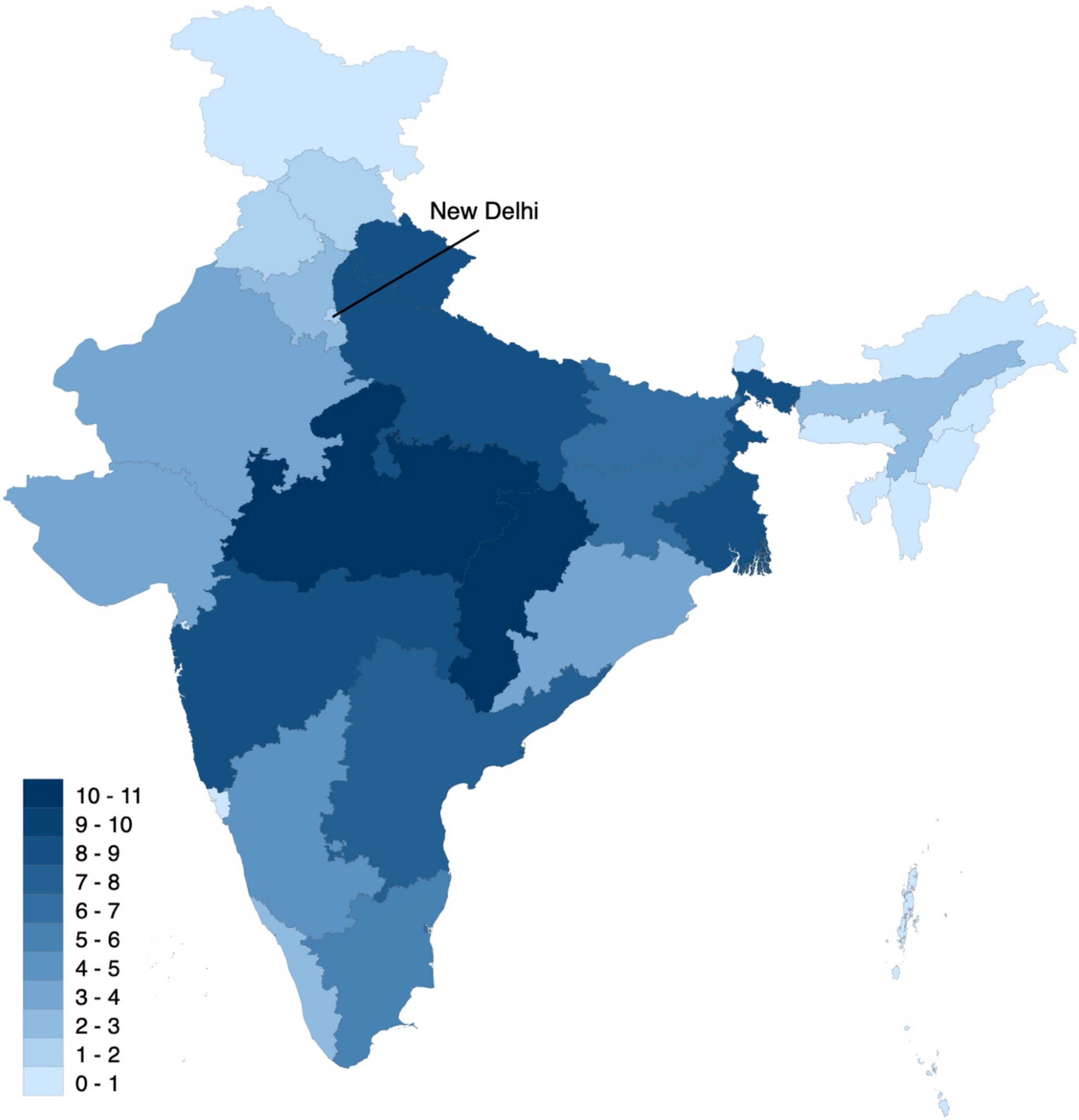

*Notes:* The figure presents the state-level variation in exposure to the forced sterilization policy as measured by the number of sterilizations performed in 1976–77 (expressed in 100,000s). Darker shades denote a greater number of sterilizations performed.

Figure 3: Households Who Use Public Health Care Facilities (in Percentages)

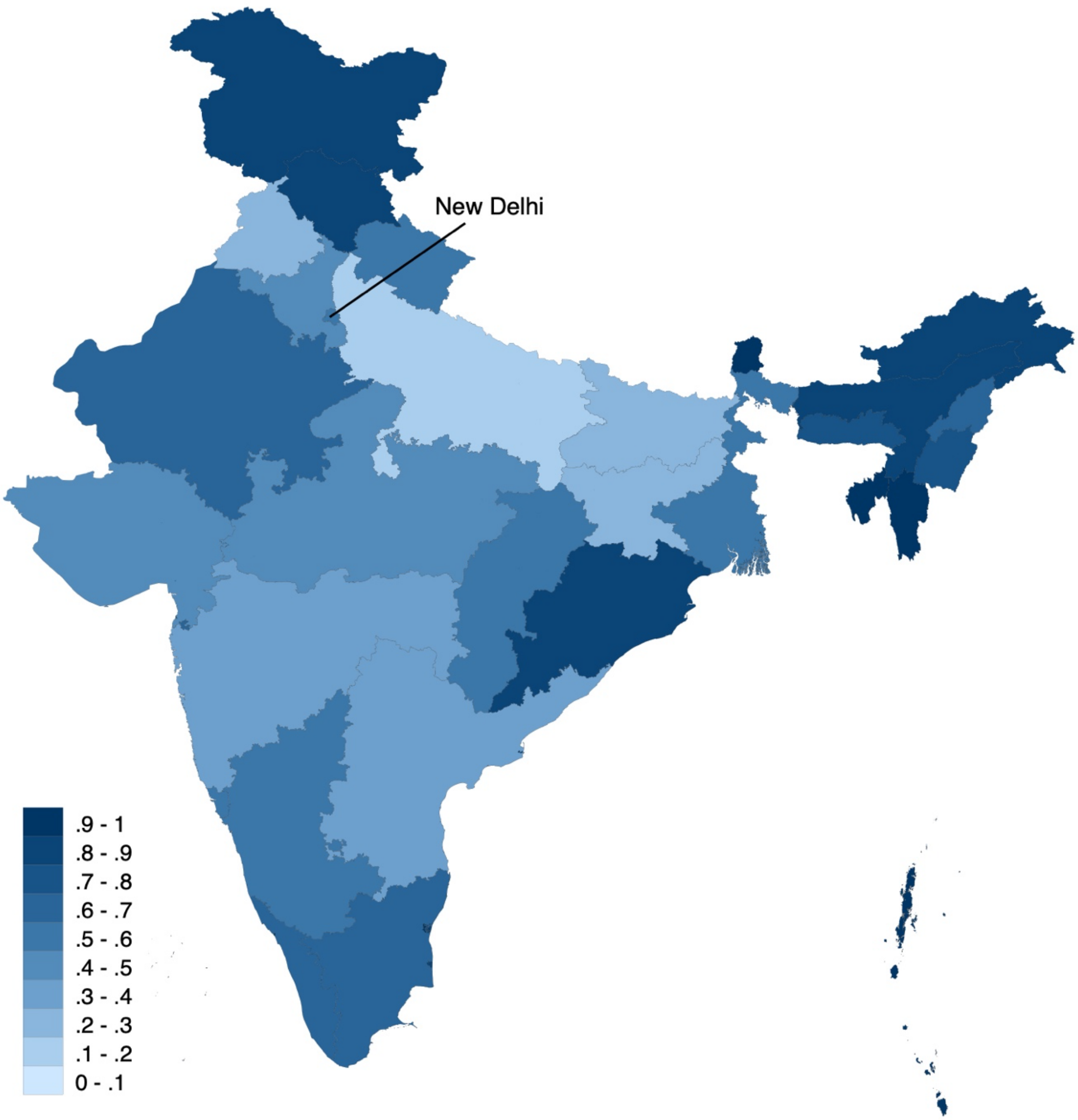

*Notes:* The figure presents the state-level variation in the usage of public health care facilities in India (expressed in percentages). Darker shades denote a higher share of public health care facility use. The dataset on public health care facility use is available at a more granular level (such as at the district and NFHS-4 cluster level). It is grouped at the state-level for ease of visualization and comparison with the sterilization figure (Figure 1).

Figure 4: Association Between Number of Sterilizations in 1976-77 and Public Health Care Use

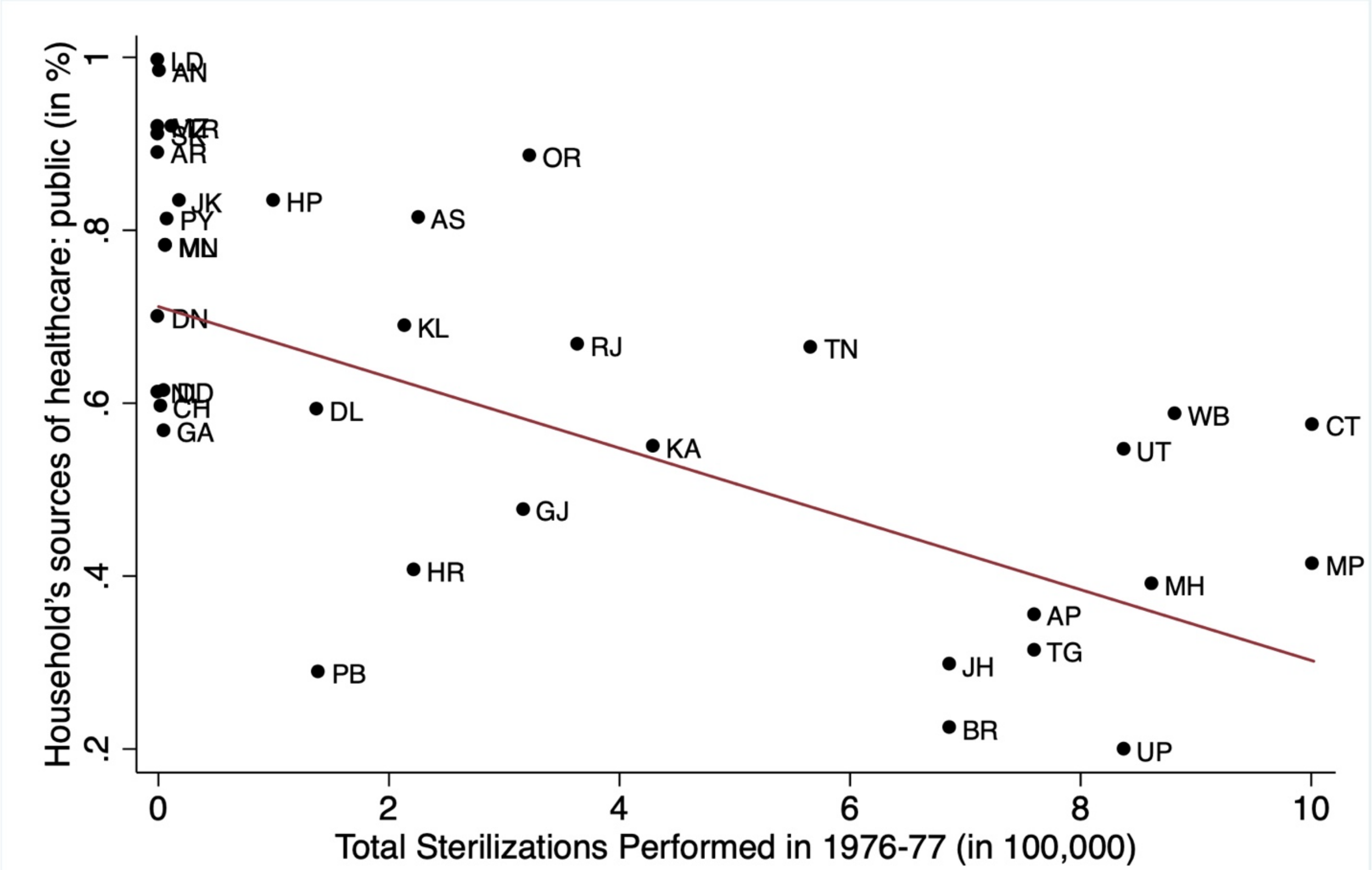

*Notes:* The figure presents the correlation plot of the state-level total number of sterilizations performed in 1976–77 (expressed in 100,000s) and the household's usage of public health care facilities in India in 2015–16 (expressed in percentages). The fitted lines are weighted by the population of the state and union territory.

Figure 5: Exogeneity of the Instrument

Panel A: Association between distance from New Delhi to state capitals and excess sterilization in 1975-76 ***(previous year)***

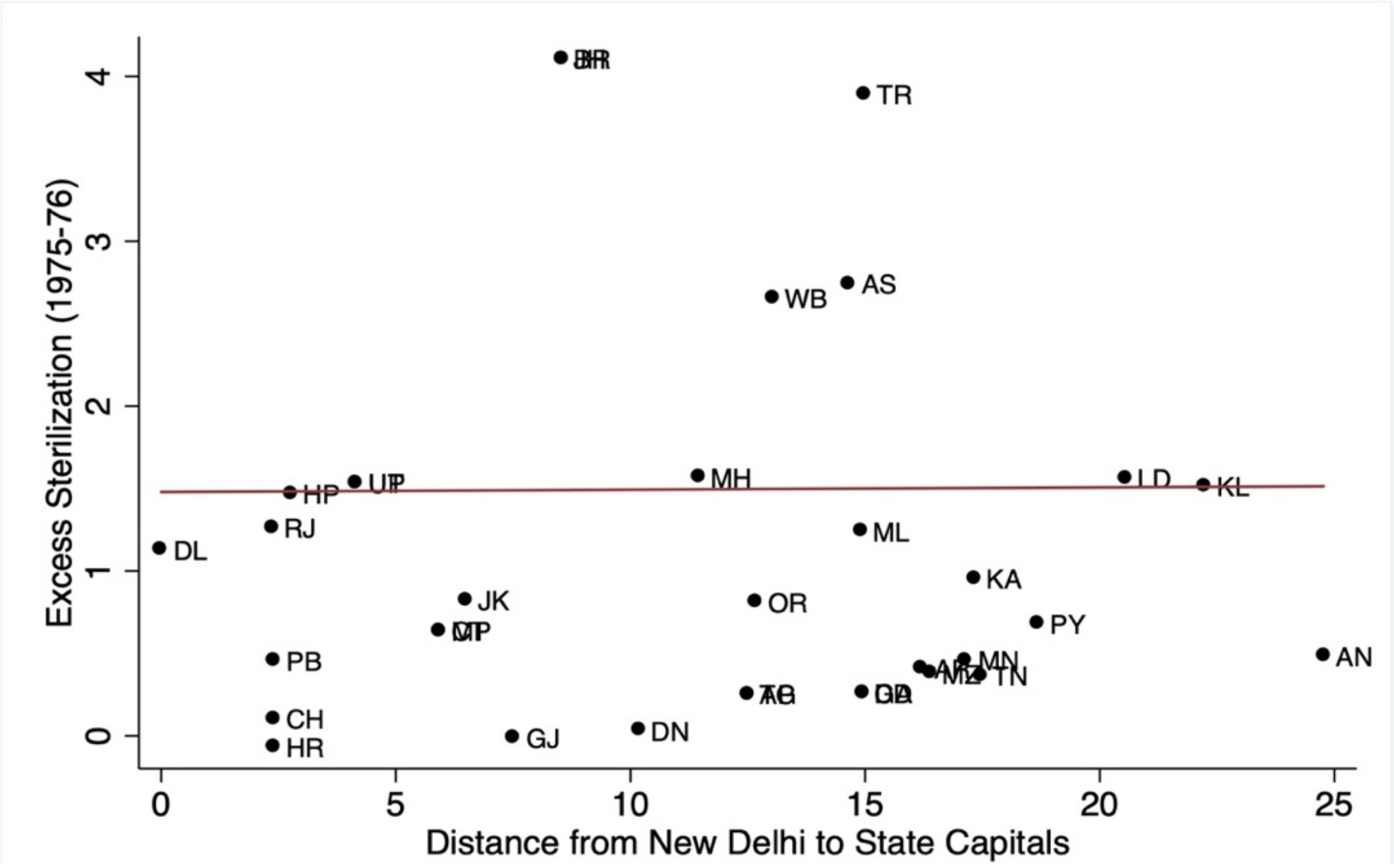

Panel B: Association between distance from New Delhi to state capitals and excess tubectomy

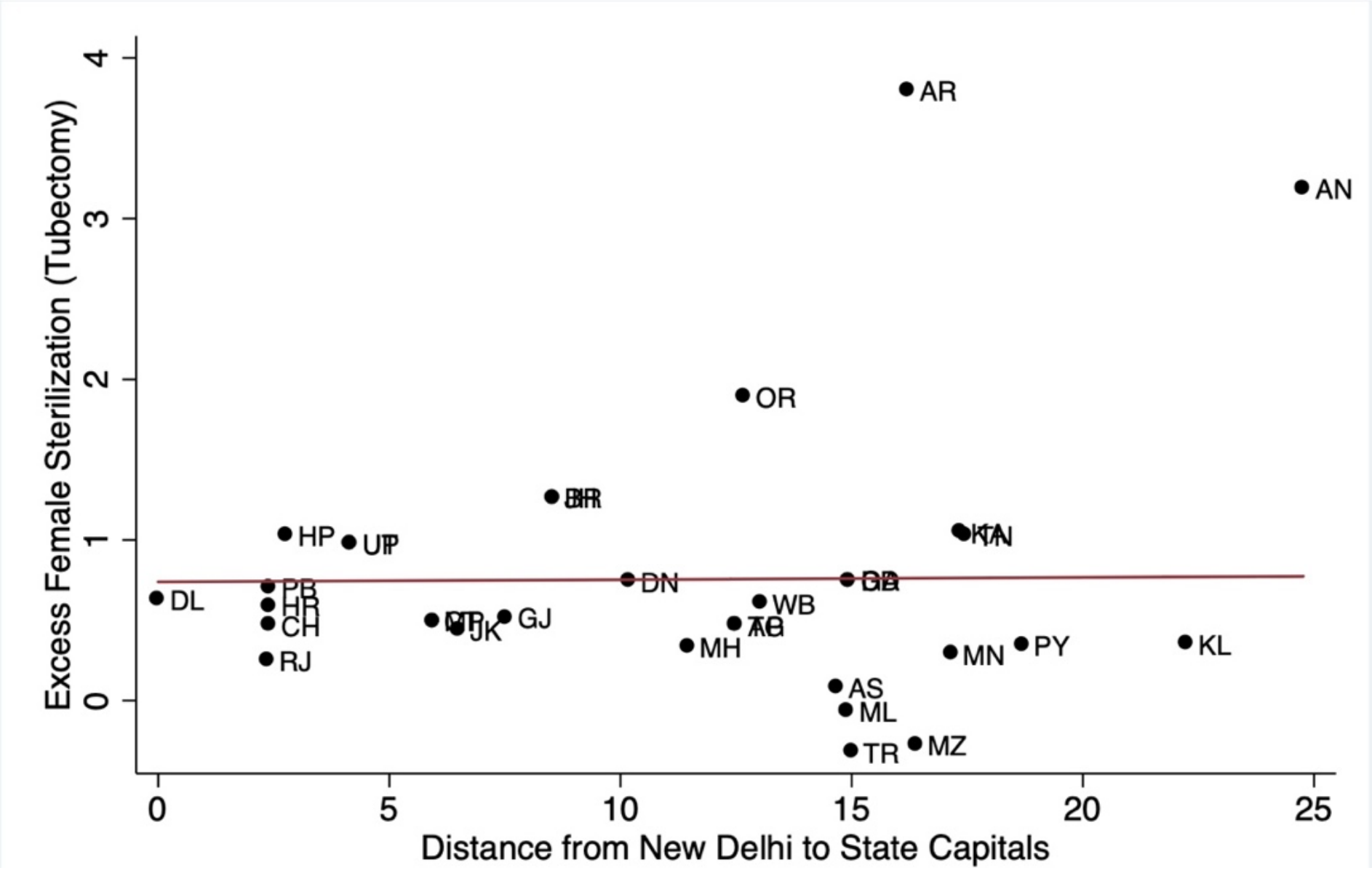

*Notes:* The figure presents the exogeneity of the instrument. Panel A presents the correlation between state-level excess sterilizations performed in 1975–76 (*previous year*) and the distance from New Delhi to state capitals (expressed in 100 kilometers). Panel B presents the correlation between state-level excess Tubectomy performed in 1976–77 and the distance from New Delhi to state capitals. The fitted lines are weighted by the population of the state and union territory.

Figure 6: Falsification Exercise to Test the Threats against Instrument Validity

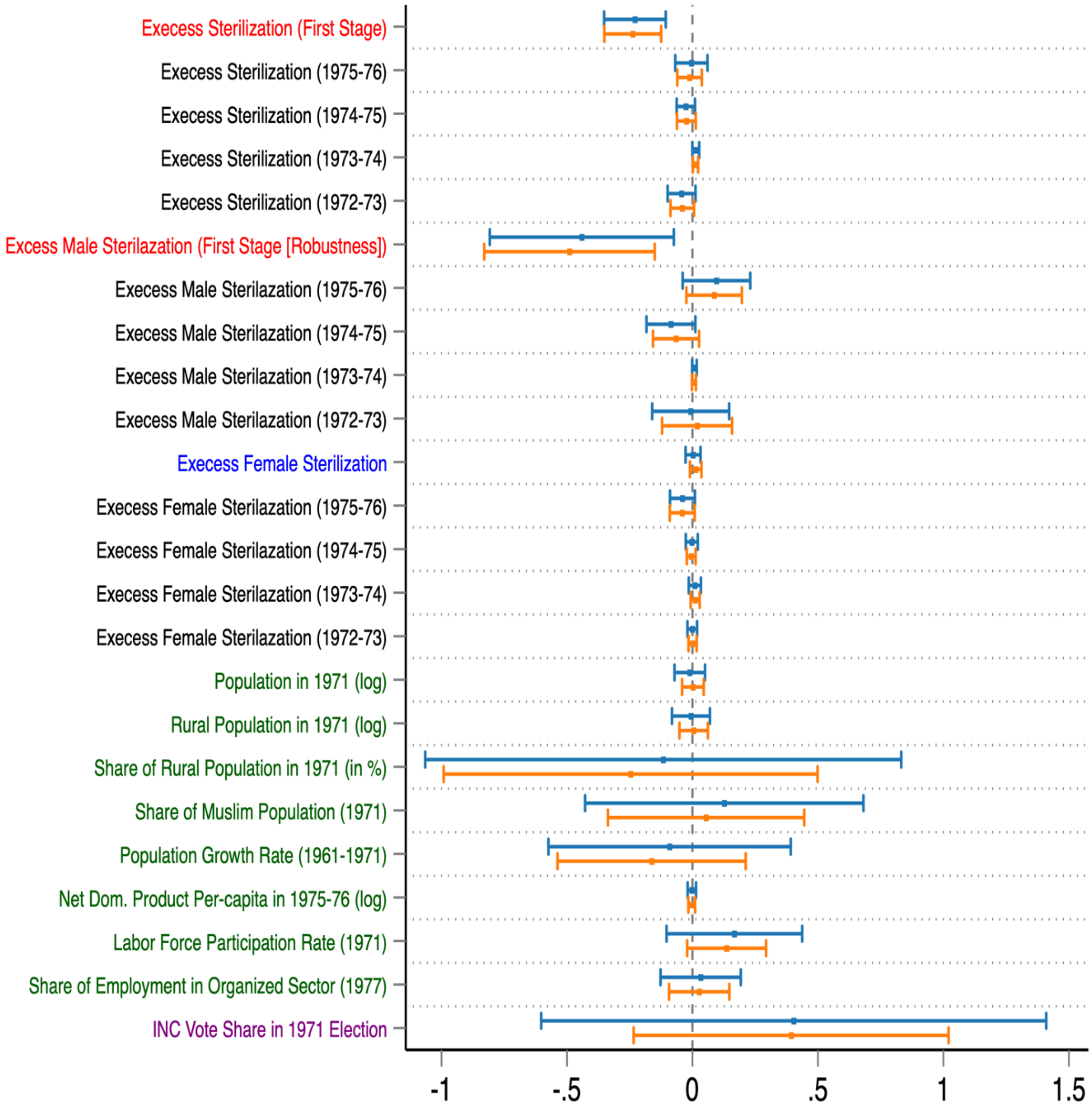

*Notes:* The estimates correspond to the specifications from column 1 (top-blue) and column 4 (bottom-orange) in Table 2. Please see notes to Table 2. Each estimate comes from a separate regression. The explanatory variable is the distance from New Delhi to state capitals (expressed in 100 kilometers). The dots are the estimated coefficients and the horizontal lines represent the 95 percent confidence intervals. Share of employment in organized sector is only available since 1977.

Figure 7: Correlation Plot: Confidence in Hospitals and Doctors

Panel A: Association between excess sterilizations in 1976–77 on confidence in government hospitals and doctors

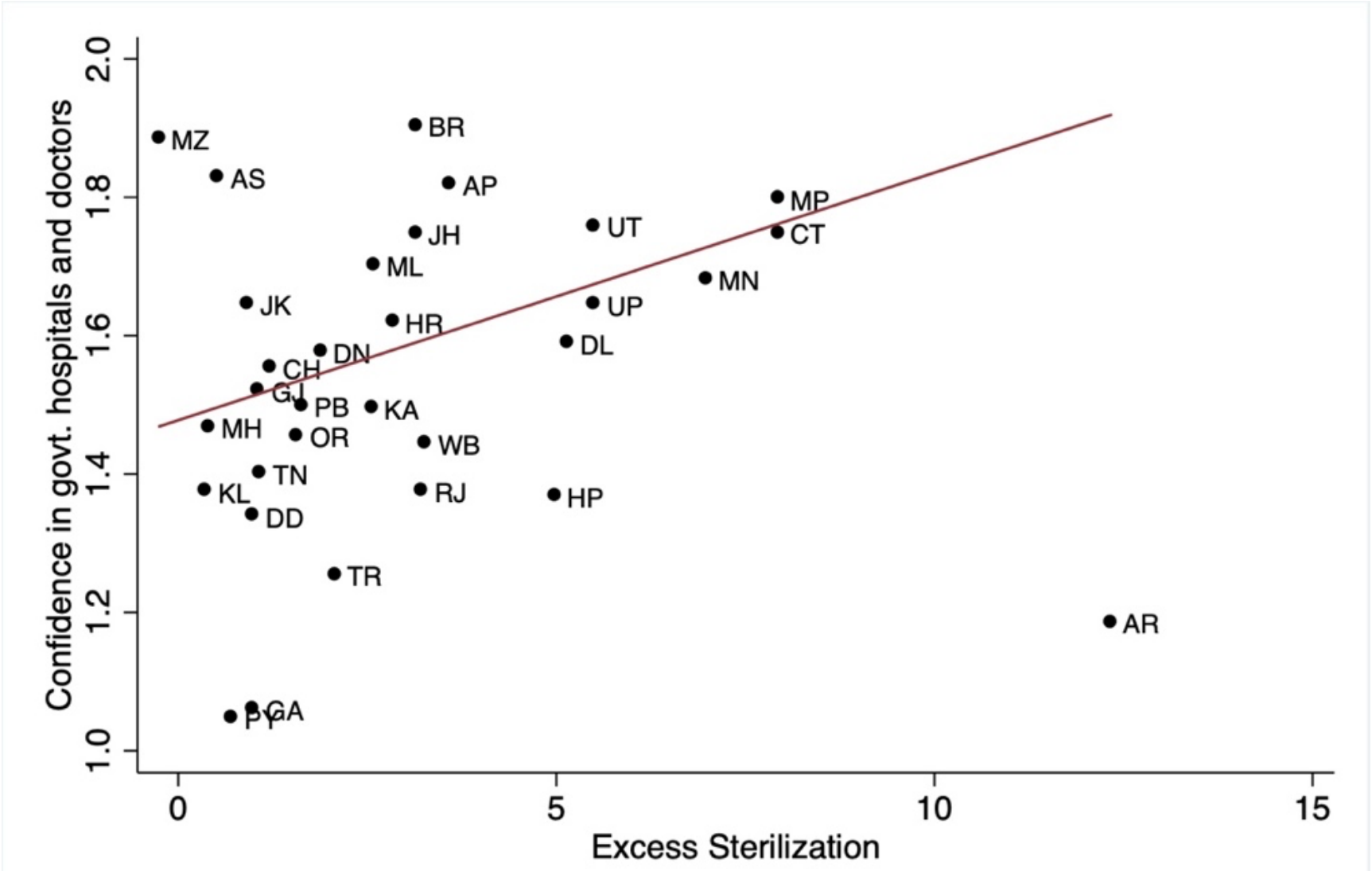

Panel B: Association between excess sterilizations in 1976–77 on confidence in private hospitals and doctors

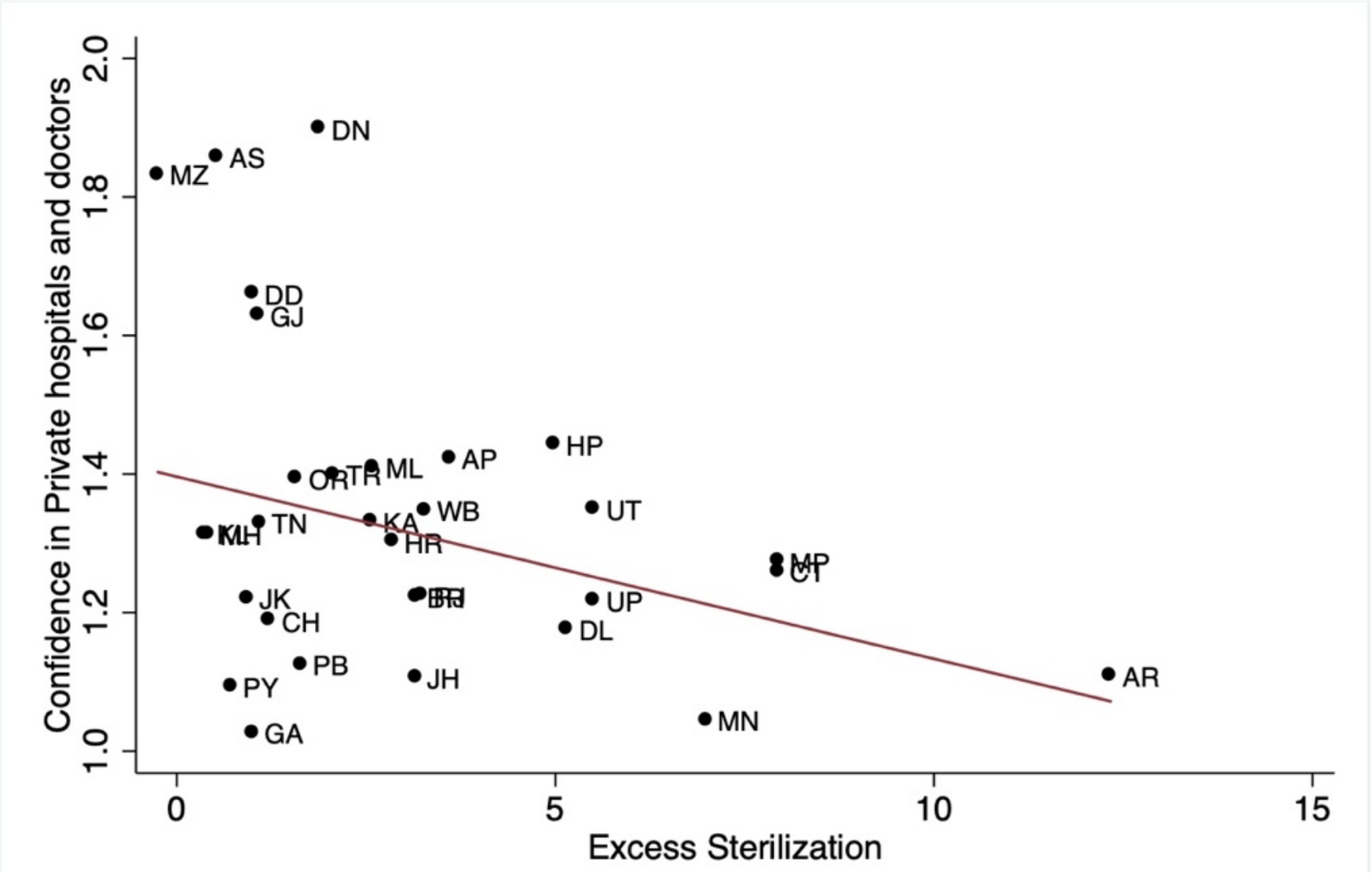

*Notes:* The figure presents the correlation plots of the mechanism. The data on confidence in hospital and doctors are from IHDS-II. It assigns the value 1 to "a great deal of confidence," 2 to "only some confidence," and 3 to "hardly any confidence at all." Therefore, higher score constitutes a lower level of confidence. Panel A plots the correlation between excess sterilizations in 1976–77 and confidence in government hospitals and doctors. Panel B plots the correlation between excess sterilizations in 1976–77 and confidence in private hospitals and doctors. The fitted lines are weighted by the population of the state and union territory.

Table 1: OLS Estimates

| | Dependent Variable: Source of Health Care - Public Sector | | | | |
|---|---|---|---|---|---|
| | (1) | (2) | (3) | (4) | (5) |
| Total Sterilizations Performed in 1976-77 (in 100,000) | -0.0370 | | | | |
| | (0.0112) | | | | |
| Total Sterilizations Performed in 1976-77 (in log) | | -0.0864 | | | |
| | | (0.0232) | | | |
| Excess Sterilization Performed in 1976-77 (in 100,000) | | | -0.0401 | | |
| | | | (0.0119) | | |
| Excess Sterilization Performed in 1976-77 (in log) | | | | -0.111 | |
| | | | | (0.0245) | |
| Excess Sterilization | | | | | -0.0367 |
| | | | | | (0.0160) |
| Household Controls | YES | YES | YES | YES | YES |
| Geographic Controls | YES | YES | YES | YES | YES |
| Health Facility Controls | YES | YES | YES | YES | YES |
| Observations | 574,022 | 574,022 | 558,755 | 547,495 | 558,755 |
| Mean of dependent variable | 0.443 | 0.443 | 0.442 | 0.442 | 0.442 |

*Notes:* Data are from India's National Family and Health Survey 2015-16 (NFHS-4). The Unit of observation is a household. Household controls include age and sex of the household head, household size, nine religion fixed effects, four caste fixed effects, 21 education of the household head fixed effects, four household wealth index fixed effects, an indicator for whether any household member is covered by health insurance, and an indicator for whether the household has a BPL card. The geographic controls include altitude of the cluster in meters, altitude squared, state level population density (in log) and an indicator whether the place of residence is urban. Health facility controls include hospital per 1000 population and doctors per 1000 population at the state level. Robust standard errors in parentheses clustered at the state level.

Table 2: Instrumental Variable Estimates

| | (1) | (2) | (3) | (4) |
|---|---|---|---|---|
| | **Panel A: First Stage Estimates** Dependent variable: Excess Sterilization | | | |
| Distance from New Delhi to State Capitals (in 100km) | -0.229 | -0.249 | -0.245 | -0.238 |
| | (0.060) | (0.059) | (0.058) | (0.056) |
| Mean of dependent variable | 3.150 | 3.177 | 3.177 | 3.177 |
| F Statistics of Excluded Instrument | 14.42 | 17.88 | 17.80 | 18.27 |
| | **Panel B: Second Stage Estimates** Dependent variable: Source of Health Care - Public Sector | | | |
| Excess Sterilization | -0.071 | -0.056 | -0.057 | -0.059 |
| | (0.025) | (0.024) | (0.024) | (0.020) |
| Mean of dependent variable | 0.448 | 0.443 | 0.442 | 0.442 |
| | **Panel C: Reduced Form Estimates** Dependent variable: Source of Health Care - Public Sector | | | |
| Distance from New Delhi to State Capitals (in 100km) | 0.016 | 0.014 | 0.014 | 0.014 |
| | (0.007) | (0.007) | (0.007) | (0.006) |
| Household Controls | NO | YES | YES | YES |
| Geographic Controls | NO | NO | YES | YES |
| Health Facility Controls | NO | NO | NO | YES |
| Observations | 585,634 | 559,899 | 558,755 | 558,755 |

*Notes:* Data are from India's National Family and Health Survey 2015-16 (NFHS-4). The Unit of observation is a household. Household controls include age and sex of the household head, household size, nine religion fixed effects, four caste fixed effects, 21 education of the household head fixed effects, four household wealth index fixed effects, an indicator for whether any household member is covered by health insurance, and an indicator for whether the household has a BPL card. The geographic controls include altitude of the cluster in meters, altitude squared, state level population density (in log) and an indicator whether the place of residence is urban. Health facility controls include hospital per 1000 population and doctors per 1000 population at the state level. Robust standard errors in parentheses clustered at the state level.

Table 3: Alternative Measures of Variation in Forced Sterilization

| | Dependent Variable: Source of Health Care - Public Sector | | | | | | | |
|---|---|---|---|---|---|---|---|---|
| | (1) | (2) | (3) | (4) | (5) | (6) | (7) | (8) |
| INC's Vote Share in 1977 Election | 0.560 | 0.440 | 0.333 | 0.384 | | | | |
| | (0.0643) | (0.0623) | (0.0710) | (0.0615) | | | | |
| Change in INC's Vote Share (1977-1971) | | | | | 0.535 | 0.405 | 0.363 | 0.398 |
| | | | | | (0.0757) | (0.0739) | (0.0664) | (0.0582) |
| Individual Controls | NO | YES | YES | YES | NO | YES | YES | YES |
| Household Controls | NO | NO | YES | YES | NO | NO | YES | YES |
| Geographic Controls | NO | NO | NO | YES | NO | NO | NO | YES |
| Observations | 456,719 | 437,530 | 437,530 | 437,530 | 349,030 | 332,703 | 332,703 | 332,703 |
| Mean of dependent variable | 0.436 | 0.430 | 0.430 | 0.430 | 0.435 | 0.429 | 0.429 | 0.429 |
| Mean of INC's Vote Share in 1977 Election | 0.384 | 0.382 | 0.382 | 0.382 | | | | |
| Mean of Change in INC's Vote Share (1977-1971) | | | | | -0.158 | -0.160 | -0.160 | -0.160 |

*Notes:* Data are from India's National Family and Health Survey 2015-16 (NFHS-4). The Unit of observation is a household. INC's Vote Share in 1977 Election measures the constituency level variation in INC candidates' vote share in the 1977 election (a proxy measure of excess sterilization). Change in INC's Vote Share (1977-1971) measures the constituency level change in INC candidates' vote share in 1977 compared with the 1971 election (as a second proxy measure of excess sterilization). Household controls include age and sex of the household head, household size, nine religion fixed effects, four caste fixed effects, 21 education of the household head fixed effects, four household wealth index fixed effects, an indicator for whether any household member is covered by health insurance, and an indicator for whether the household has a BPL card. The geographic controls include altitude of the cluster in meters, altitude squared, state level population density (in log) and an indicator whether the place of residence is urban. Health facility controls include hospital per 1000 population and doctors per 1000 population at the state level. Robust standard errors in parentheses clustered at the parliament constituency level.

Table 4: Mechanism: Reasons

| Dependent variable: | No nearby facility | Facility timing not convenient | Health personnel often absent | Waiting time too long | Poor quality of care | Other |
|---|---|---|---|---|---|---|
| | (1) | (2) | (3) | (4) | (5) | (6) |
| Excess Sterilization | -0.00161 | -0.0193 | -0.00404 | 0.00663 | 0.0596 | 0.00805 |
| | (0.00720) | (0.0103) | (0.00883) | (0.00923) | (0.0163) | (0.00352) |
| Household Controls | YES | YES | YES | YES | YES | YES |
| Geographic Controls | YES | YES | YES | YES | YES | YES |
| Health Facility Controls | YES | YES | YES | YES | YES | YES |
| Observations | 274,693 | 274,693 | 274,693 | 274,693 | 274,693 | 274,693 |
| Mean of dependent variable | 0.445 | 0.263 | 0.149 | 0.408 | 0.483 | 0.0440 |

*Notes:* Data are from India's National Family and Health Survey 2015-16 (NFHS-4). The Unit of observation is a household. Household controls include age and sex of the household head, household size, nine religion fixed effects, four caste fixed effects, 21 education of the household head fixed effects, four household wealth index fixed effects, an indicator for whether any household member is covered by health insurance, and an indicator for whether the household has a BPL card. The geographic controls include altitude of the cluster in meters, altitude squared, state level population density (in log) and an indicator whether the place of residence is urban. Health facility controls include hospital per 1000 population and doctors per 1000 population at the state level. Robust standard errors in parentheses clustered at the state level.

Table 5: Mechanism: Confidence in Institutions

| Dependent variable: | Confidence: Government hospitals and doctors<br>(1) | Confidence: Private hospitals and doctors<br>(2) |
|---|---|---|
| Excess Sterilization | 0.0605 | -0.0325 |
| | (0.0178) | (0.0167) |
| Household Controls | YES | YES |
| Geographic Controls | YES | YES |
| Health Facility Controls | YES | YES |
| Observations | 40,562 | 40,549 |
| Mean of dependent variable | 1.577 | 1.308 |

*Notes:* The data on confidence in hospital and doctors are from IHDS-II. The unit of observation is a household. The IDHS-II assigns the value 1 to "a great deal of confidence," 2 to "only some confidence," and 3 to "hardly any confidence at all." Therefore, higher score constitutes a lower level of confidence. The household controls include household size, income, ten source of main income fixed effects, eight religion fixed effects, five caste fixed effects, two wealth class fixed effects (poor, middle class, (comfortable as reference group)), 16 education of the household head fixed effects, an indicator for whether any household member is covered by government health insurance, an indicator for whether any household member is covered by private health insurance, and an indicator for whether the household has a BPL card. The geographic controls include state level population density (in log) and three place of residence fixed effects. Health facility controls include hospital per 1000 population and doctors per 1000 population at the state level. Robust standard errors in parentheses clustered at the state level.

# Appendix for Online Publication

# Understanding the Paradox of Primary Health Care Use: Empirical Evidence from India

Pramod Kumar Sur
*Asian Growth Research Institute (AGI) and Osaka University*
pramodsur@gmail.com

## Section A: Figures

Figure A1: Private Health Care Spending around the World

Panel A: Share of Out-of-Pocket Expenditure on Health Care

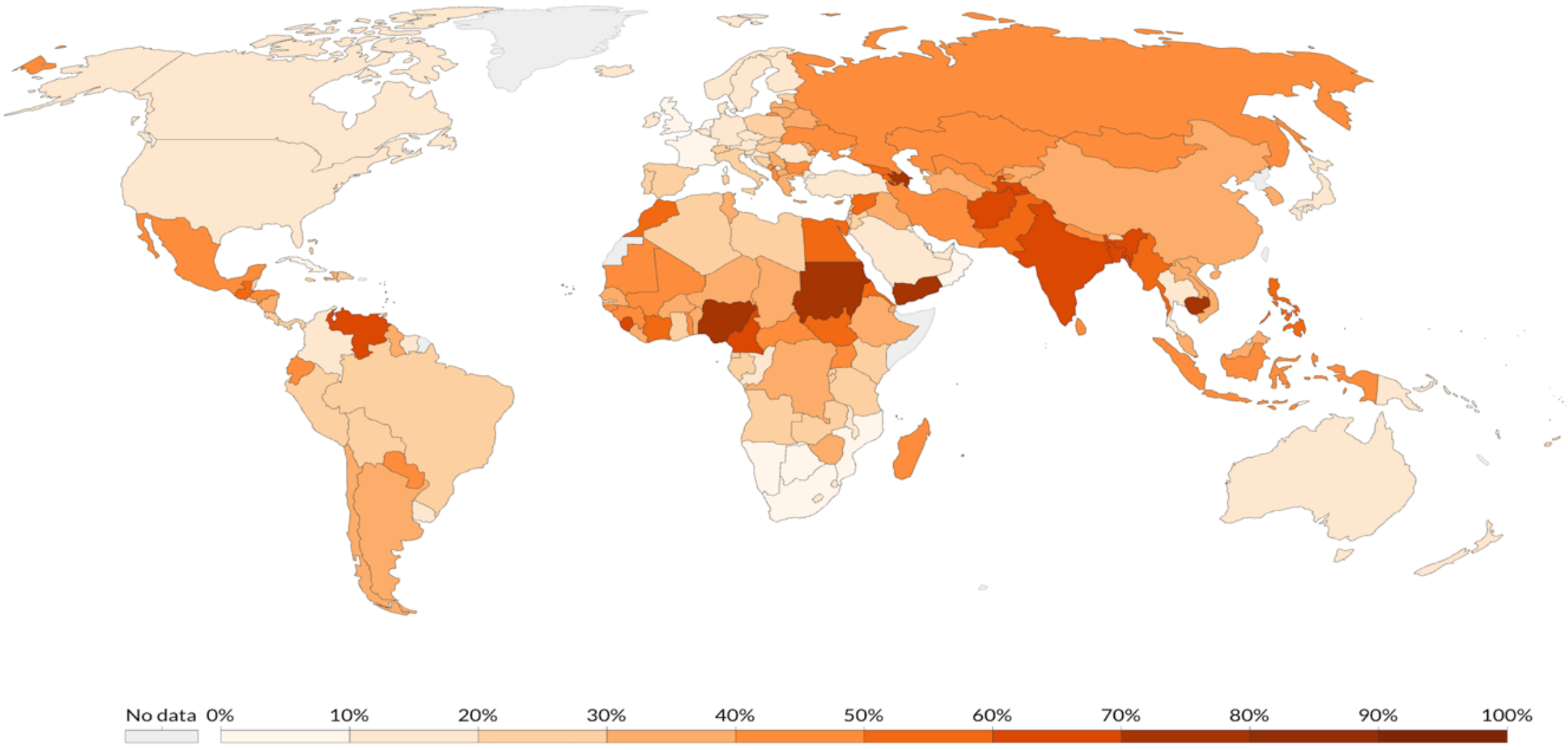

Panel B: Composition of Health Care Spending in Lower Income Countries

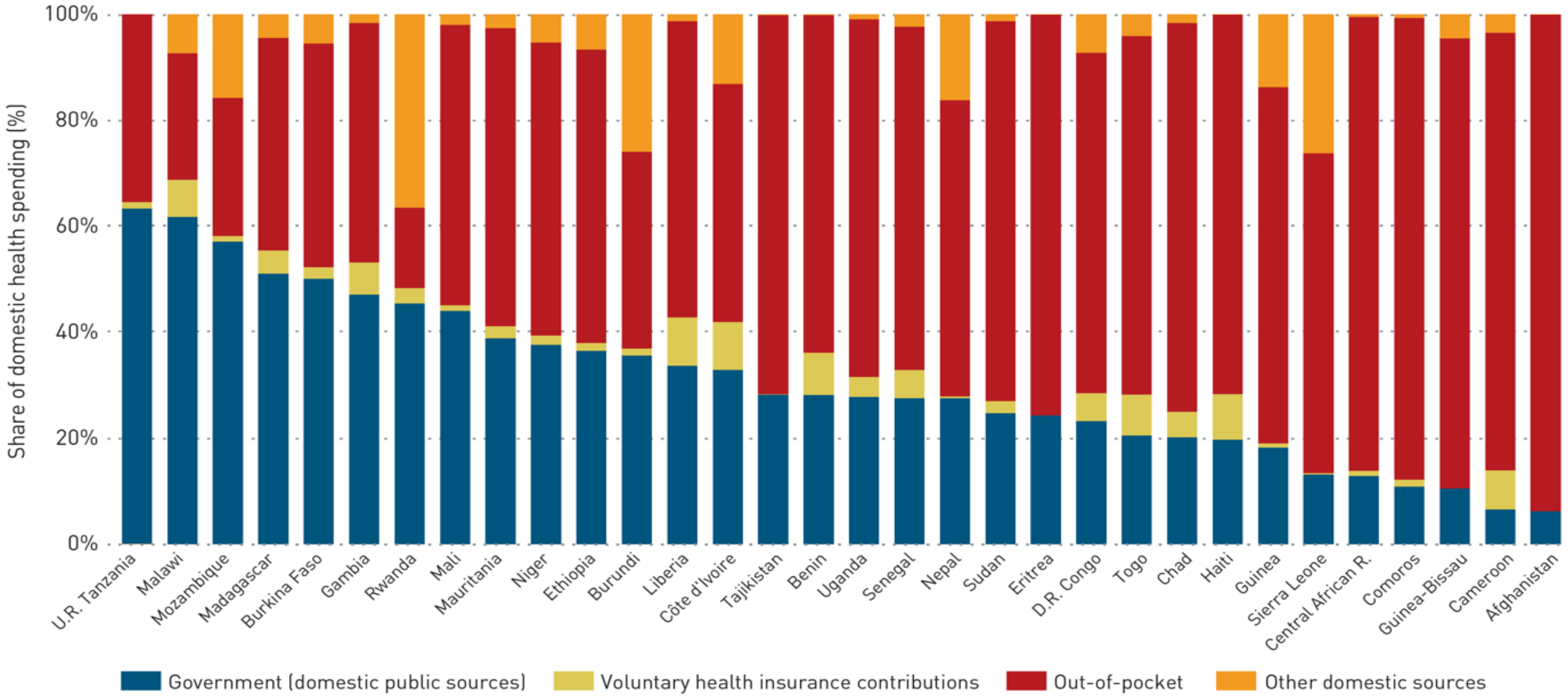

*Notes:* Panel A of Figure A1 presents the country-level share of out-of-pocket expenditure on health care as percentage of total healthcare expenditure in 2014. Panel B presents the composition of domestic health care expenditure by main financing sources in 32 lower income countries in 2018. Out-of-pocket in Panel A refers to direct outlays made by households, including gratuities and in-kind payments, to health care providers.

Figure Source: Panel A: Our World in Data.

Panel B: (WHO 2020) https://apps.who.int/iris/handle/10665/337859

Figure A2: Number of Sterilizations Performed in India (1956-82)

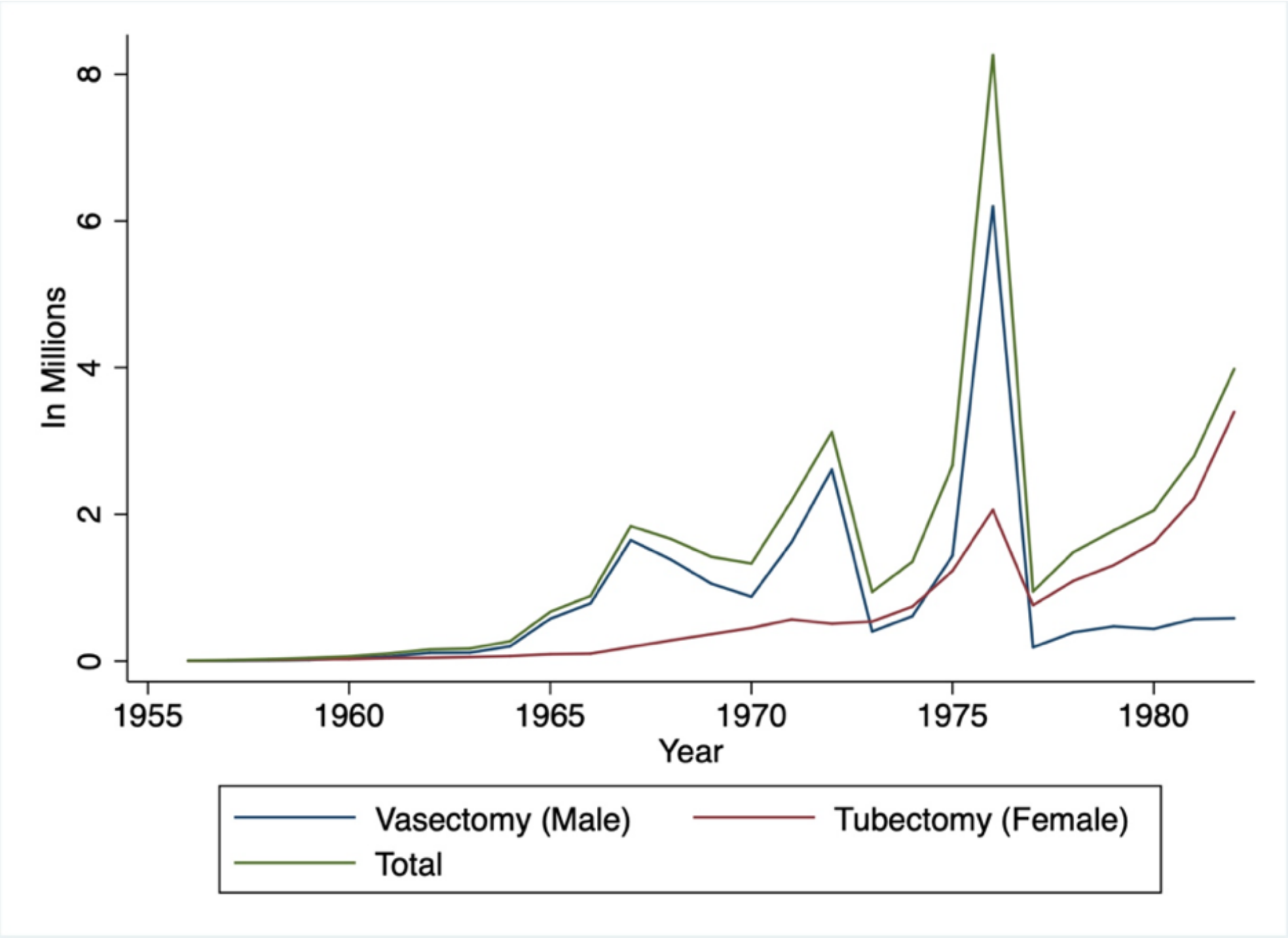

*Notes:* Figure A2 presents the total number of sterilizations along with the types of sterilization performed in India every year since the beginning of the program in 1956. The green line represents the total number of sterilizations performed every year. The blue and red lines represent the total number of vasectomies and tubectomies performed every year, respectively.

Figure A3: Geographical Distribution of NFHS-4 Clusters Matched with Assembly Constituencies (Pre-delimited)

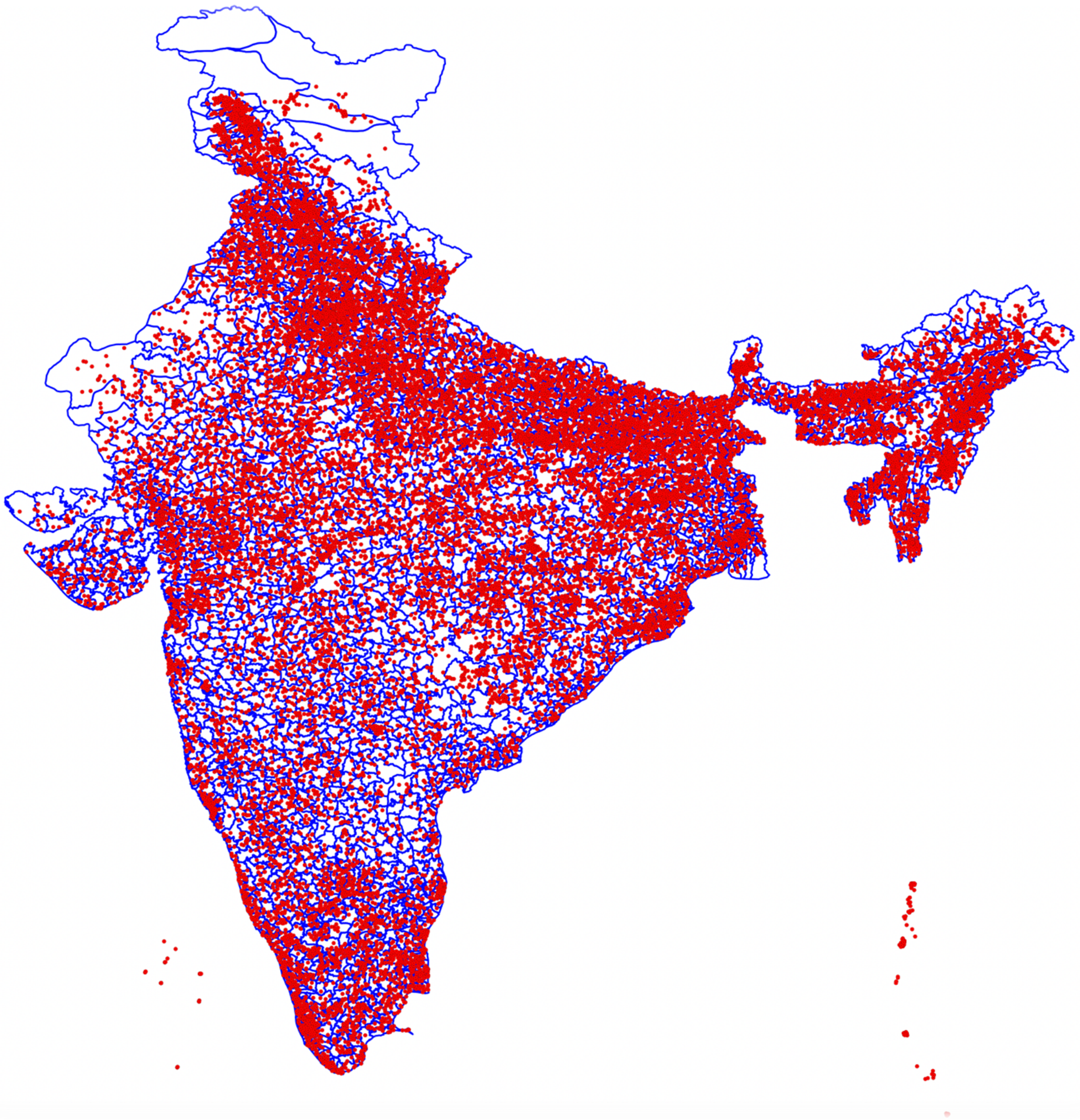

*Notes:* Figure A3 presents the geographical distribution of NFHS-4 clusters matched with assembly constituencies (pre-delimited) in India. The red circles represent the NFHS-4 clusters. The blue lines represent the borders of assembly constituencies.

Figure A4: Correlation Plot (Population Scale)

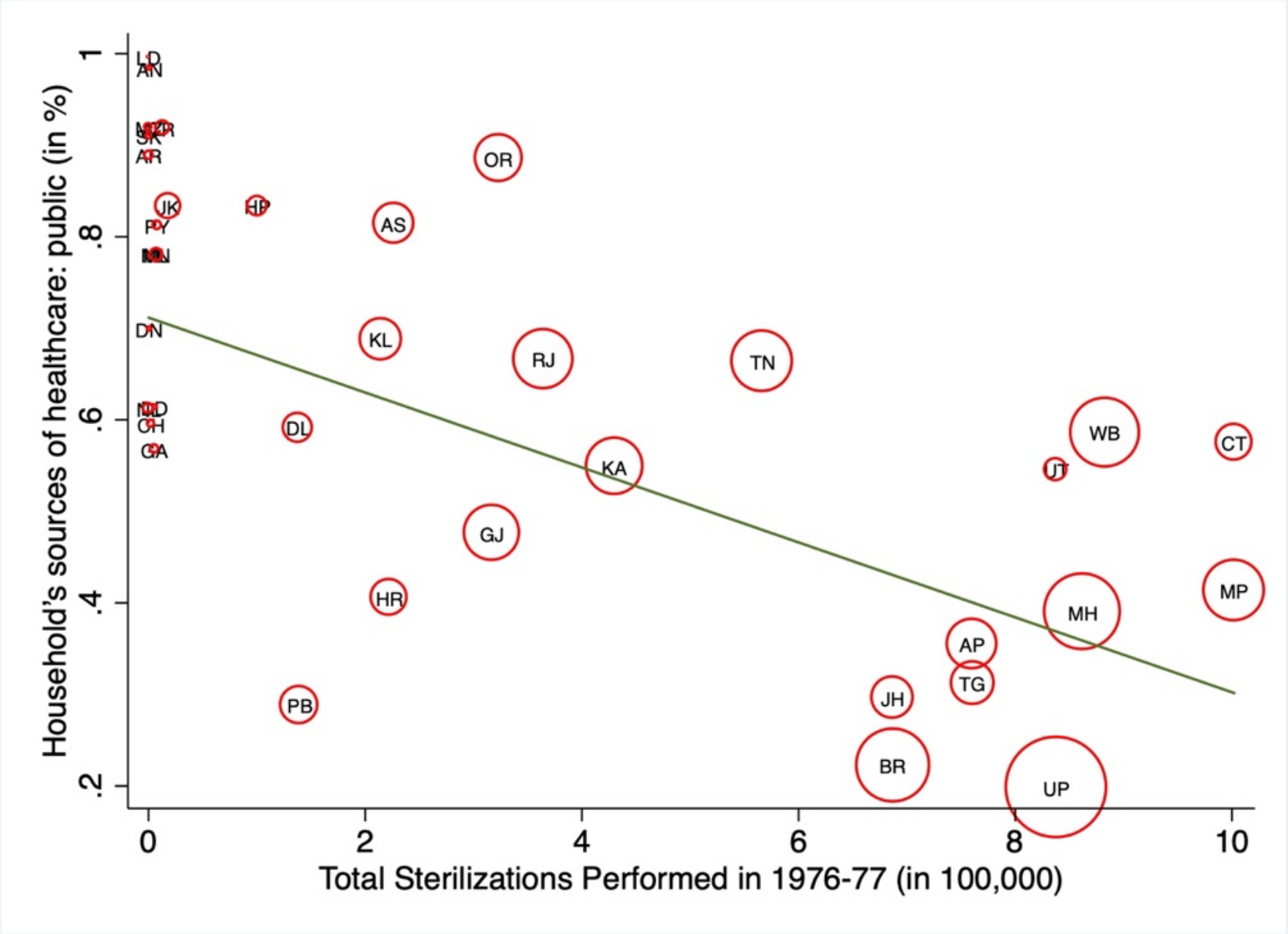

*Notes:* Figure A4 presents the correlation plot of the state-level total number of sterilizations performed in 1976–77 (expressed in 100,000s) and the household's usage of public health care facilities in India in 2015–16 (expressed in percentages). The symbols are scaled by the size of the population of the state (from the 2011 census). The fitted lines are weighted by the population of the state and union territory.

Figure A5: Number of Sterilizations Performed in 1976-77

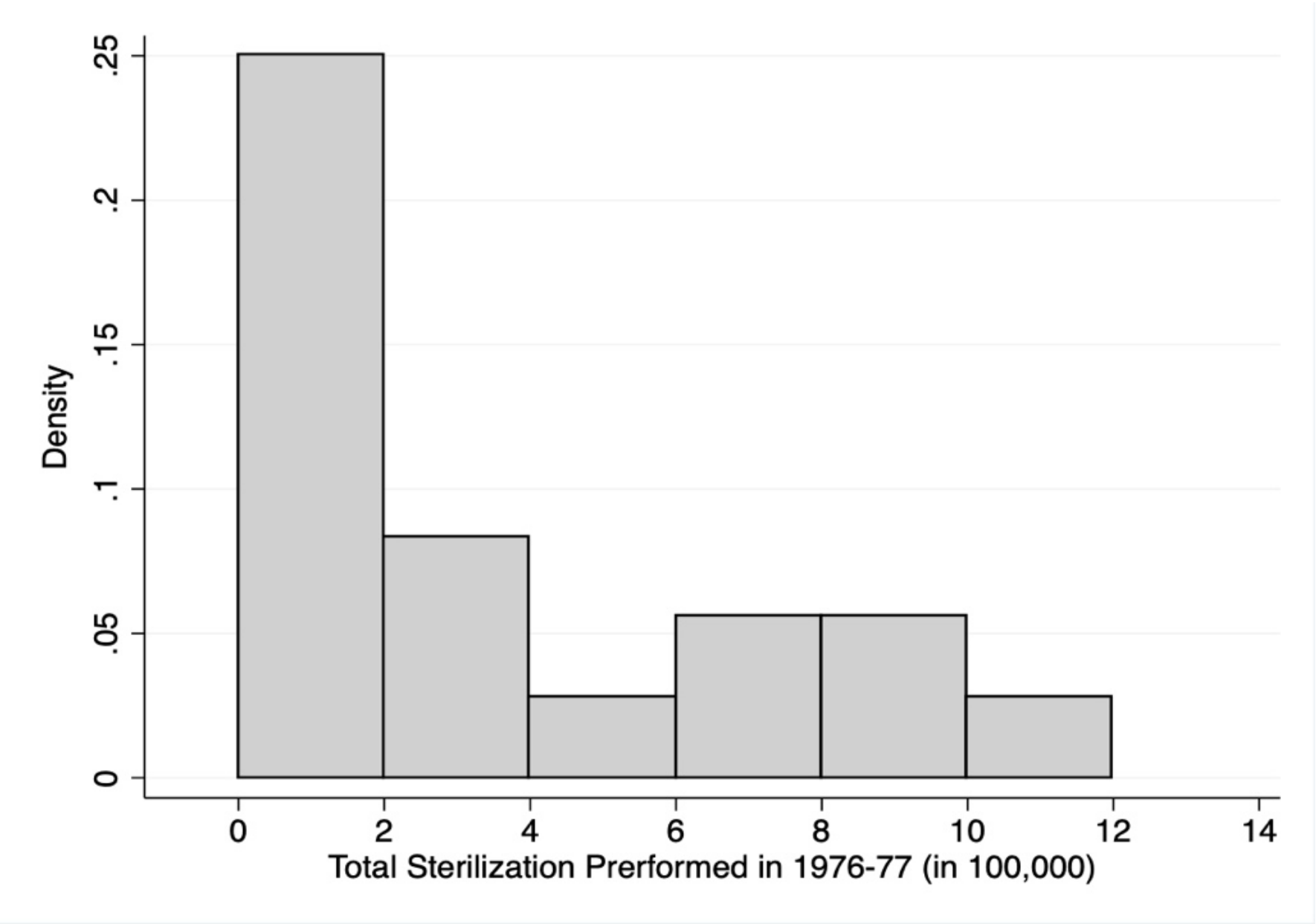

*Notes:* Figure A5 plots the distribution of the number of sterilizations performed in 1976–77 (expressed in 100,000s) at the state level.

Figure A6: Alternative Measures of Variation in Forced Sterilization

Panel A: Association between excess sterilization in 1976-77 and INC's vote share in 1977 election

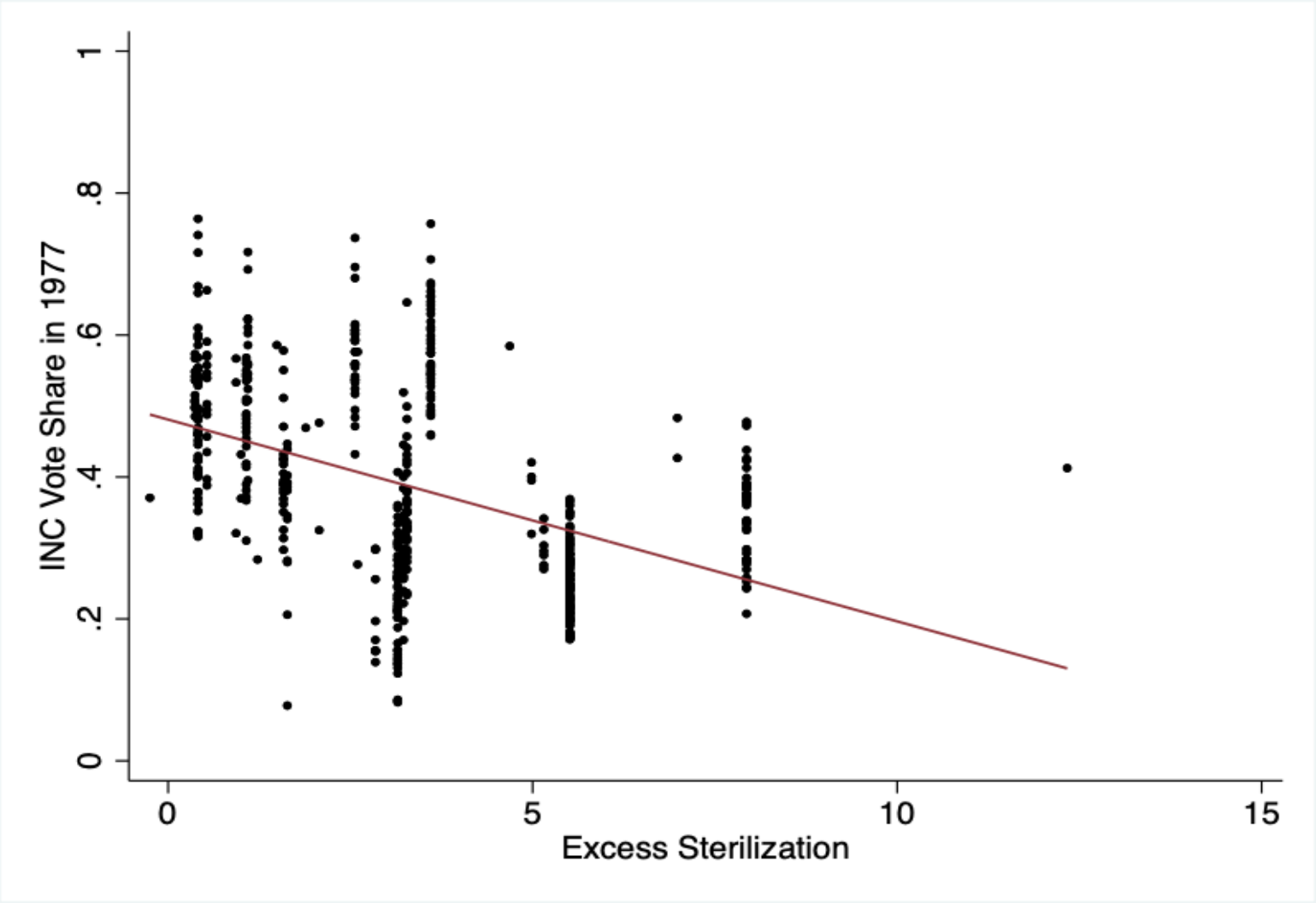

Panel B: Association between excess sterilization in 1976-77 and INC's vote share in 1971 election ***(previous election to 1977)***

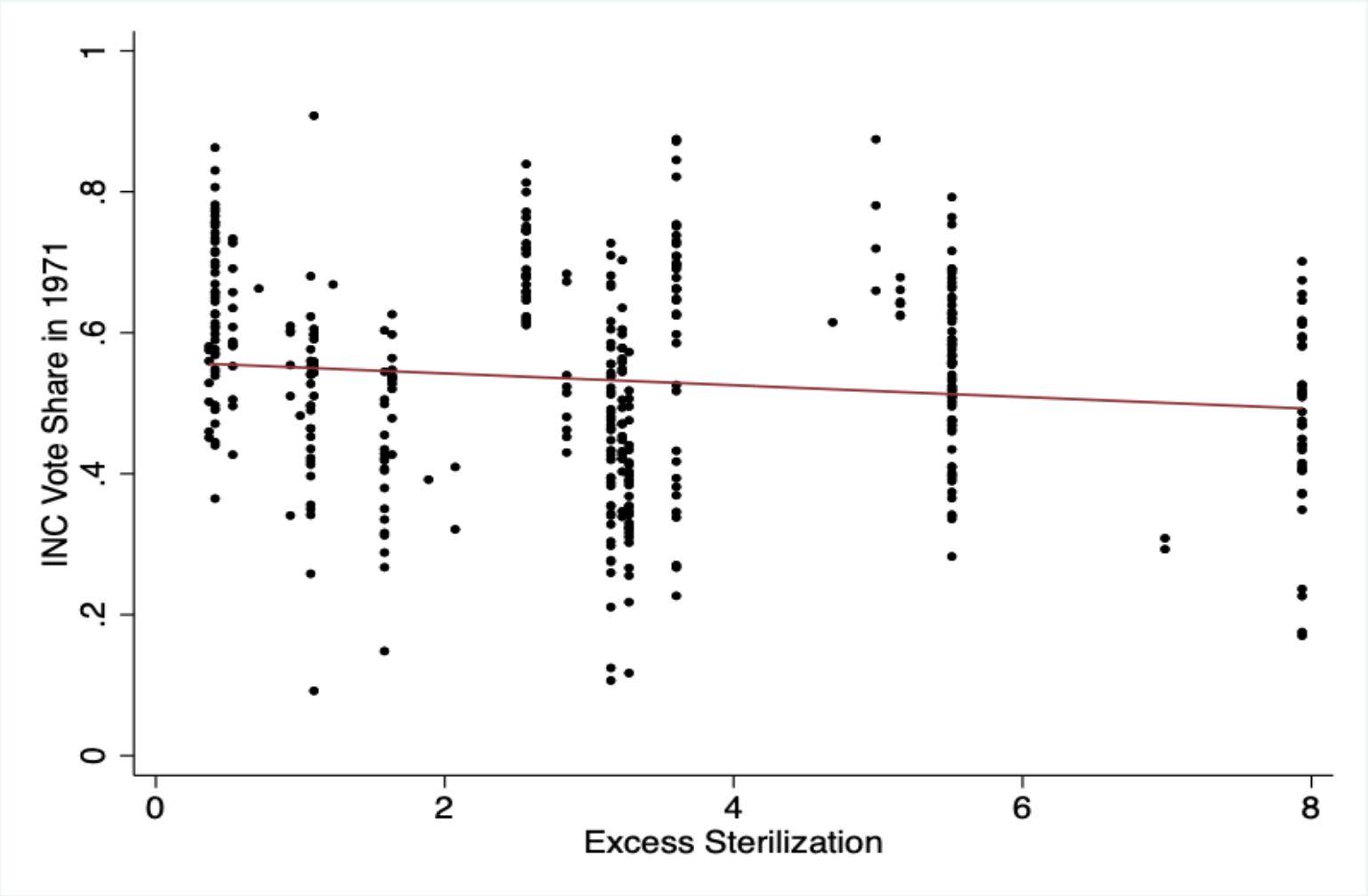

*Notes:* Figure A6 presents an alternative source of variation in forced sterilization measured by the INC party's vote share. Panel A presents the association between state-level excess sterilizations performed in 1976–77 and the constituency level variation in the INC party's vote share in the 1977 election. Panel B presents the association between state-level excess sterilizations performed in 1976–77 and the constituency level variation in the INC party's vote share in 1971, the immediate election before 1977.

## Section B: Robustness to OLS Estimates

### Table B1: Alternative Measures of Force Sterilization Policy - Total Sterilizations Performed in 1976-77 (in 100,000)

| | Dependent Variable: Source of Health Care - Public Sector | | | |
|---|---|---|---|---|
| | (1) | (2) | (3) | (4) |
| Total Sterilizations Performed in 1976-77 (in 100,000) | -0.0391 | -0.0450 | -0.0448 | -0.0370 |
| | (0.0128) | (0.0100) | (0.00966) | (0.0112) |
| Household Controls | NO | YES | YES | YES |
| Geographic Controls | NO | NO | YES | YES |
| Health Facility Controls | NO | NO | NO | YES |
| Observations | 601,509 | 575,319 | 574,022 | 574,022 |
| Mean of dependent variable | 0.449 | 0.443 | 0.443 | 0.443 |

*Notes:* Data are from India's National Family and Health Survey 2015-16 (NFHS-4). The Unit of observation is a household. Household controls include age and sex of the household head, household size, nine religion fixed effects, four caste fixed effects, 21 education of the household head fixed effects, four household wealth index fixed effects, an indicator for whether any household member is covered by health insurance, and an indicator for whether the household has a BPL card. The geographic controls include altitude of the cluster in meters, altitude squared, state level population density (in log) and an indicator whether the place of residence is urban. Health facility controls include hospital per 1000 population and doctors per 1000 population at the state level. Robust standard errors in parentheses clustered at the state level.

Table B2: Alternative Measures of Force Sterilization Policy - Total Sterilizations Performed in 1976-77 (in log)

| | Dependent Variable: Source of Health Care - Public Sector | | | |
|---|---|---|---|---|
| | (1) | (2) | (3) | (4) |
| Total Sterilizations Performed in 1976-77 (in log) | -0.105 | -0.118 | -0.114 | -0.0864 |
| | (0.0283) | (0.0240) | (0.0247) | (0.0232) |
| Household Controls | NO | YES | YES | YES |
| Geographic Controls | NO | NO | YES | YES |
| Health Facility Controls | NO | NO | NO | YES |
| Observations | 601,509 | 575,319 | 574,022 | 574,022 |
| Mean of dependent variable | 0.449 | 0.443 | 0.443 | 0.443 |

*Notes:* Data are from India's National Family and Health Survey 2015-16 (NFHS-4). The Unit of observation is a household. Household controls include age and sex of the household head, household size, nine religion fixed effects, four caste fixed effects, 21 education of the household head fixed effects, four household wealth index fixed effects, an indicator for whether any household member is covered by health insurance, and an indicator for whether the household has a BPL card. The geographic controls include altitude of the cluster in meters, altitude squared, state level population density (in log) and an indicator whether the place of residence is urban. Health facility controls include hospital per 1000 population and doctors per 1000 population at the state level. Robust standard errors in parentheses clustered at the state level.

Table B3: Alternative Measures of Force Sterilization Policy - Excess Sterilization Performed in 1976-77 (in 100,000)

| | Dependent Variable: Source of Health Care - Public Sector | | | |
|---|---|---|---|---|
| | (1) | (2) | (3) | (4) |
| Excess Sterilization Performed in 1976-77 (in 100,000) | -0.0393 | -0.0458 | -0.0449 | -0.0401 |
| | (0.0152) | (0.0128) | (0.0119) | (0.0119) |
| Household Controls | NO | YES | YES | YES |
| Geographic Controls | NO | NO | YES | YES |
| Health Facility Controls | NO | NO | NO | YES |
| Observations | 585,634 | 559,899 | 558,755 | 558,755 |
| Mean of dependent variable | 0.448 | 0.443 | 0.442 | 0.442 |

*Notes:* Data are from India's National Family and Health Survey 2015-16 (NFHS-4). The Unit of observation is a household. Household controls include age and sex of the household head, household size, nine religion fixed effects, four caste fixed effects, 21 education of the household head fixed effects, four household wealth index fixed effects, an indicator for whether any household member is covered by health insurance, and an indicator for whether the household has a BPL card. The geographic controls include altitude of the cluster in meters, altitude squared, state level population density (in log) and an indicator whether the place of residence is urban. Health facility controls include hospital per 1000 population and doctors per 1000 population at the state level. Robust standard errors in parentheses clustered at the state level.

Table B4: Alternative Measures of Force Sterilization Policy - Excess Sterilization Performed in 1976-77 (in log)

| | Dependent Variable: Source of Health Care - Public Sector | | | |
|---|---|---|---|---|
| | (1) | (2) | (3) | (4) |
| Excess Sterilization Performed in 1976-77 (in log) | -0.109 | -0.129 | -0.126 | -0.111 |
| | (0.0301) | (0.0239) | (0.0239) | (0.0245) |
| Household Controls | NO | YES | YES | YES |
| Geographic Controls | NO | NO | YES | YES |
| Health Facility Controls | NO | NO | NO | YES |
| Observations | 574,237 | 548,577 | 547,495 | 547,495 |
| Mean of dependent variable | 0.448 | 0.442 | 0.442 | 0.442 |

*Notes:* Data are from India's National Family and Health Survey 2015-16 (NFHS-4). The Unit of observation is a household. Household controls include age and sex of the household head, household size, nine religion fixed effects, four caste fixed effects, 21 education of the household head fixed effects, four household wealth index fixed effects, an indicator for whether any household member is covered by health insurance, and an indicator for whether the household has a BPL card. The geographic controls include altitude of the cluster in meters, altitude squared, state level population density (in log) and an indicator whether the place of residence is urban. Health facility controls include hospital per 1000 population and doctors per 1000 population at the state level. Robust standard errors in parentheses clustered at the state level.

Table B5: Alternative Measures of Force Sterilization Policy - Excess Sterilization

| | Dependent Variable: Source of Health Care - Public Sector | | | |
|---|---|---|---|---|
| | (1) | (2) | (3) | (4) |
| Excess Sterilization | -0.0321 | -0.0316 | -0.0330 | -0.0367 |
| | (0.0192) | (0.0160) | (0.0155) | (0.0160) |
| Household Controls | NO | YES | YES | YES |
| Geographic Controls | NO | NO | YES | YES |
| Health Facility Controls | NO | NO | NO | YES |
| Observations | 585,634 | 559,899 | 558,755 | 558,755 |
| Mean of dependent variable | 0.448 | 0.443 | 0.442 | 0.442 |

*Notes:* Data are from India's National Family and Health Survey 2015-16 (NFHS-4). The Unit of observation is a household. Household controls include age and sex of the household head, household size, nine religion fixed effects, four caste fixed effects, 21 education of the household head fixed effects, four household wealth index fixed effects, an indicator for whether any household member is covered by health insurance, and an indicator for whether the household has a BPL card. The geographic controls include altitude of the cluster in meters, altitude squared, state level population density (in log) and an indicator whether the place of residence is urban. Health facility controls include hospital per 1000 population and doctors per 1000 population at the state level. Robust standard errors in parentheses clustered at the state level.

Table B6: Alternative Measures of Force Sterilization Policy - Male Sterilization

| | Dependent Variable: Source of Health Care - Public Sector | | | | |
|---|---|---|---|---|---|
| | (1) | (2) | (3) | (4) | (5) |
| Total Vasectomies Performed in 1976-77 (in 100,000) | -0.0405 | | | | |
| | (0.0129) | | | | |
| Total Vasectomies Performed in 1976-77 (in log) | | -0.0826 | | | |
| | | (0.0237) | | | |
| Excess Vasectomies Performed in 1976-77 (in 100,000) | | | -0.0391 | | |
| | | | (0.0127) | | |
| Excess Vasectomies Performed in 1976-77 (in log) | | | | -0.0978 | |
| | | | | (0.0231) | |
| Excess Male Sterilization (Vasectomy) | | | | | -0.0143 |
| | | | | | (0.00538) |
| Household Controls | YES | YES | YES | YES | YES |
| Geographic Controls | YES | YES | YES | YES | YES |
| Health Facility Controls | YES | YES | YES | YES | YES |
| Observations | 574,022 | 574,022 | 558,755 | 558,755 | 558,755 |
| Mean of dependent variable | 0.443 | 0.443 | 0.442 | 0.442 | 0.442 |

*Notes:* Data are from India's National Family and Health Survey 2015-16 (NFHS-4). The Unit of observation is a household. Household controls include age and sex of the household head, household size, nine religion fixed effects, four caste fixed effects, 21 education of the household head fixed effects, four household wealth index fixed effects, an indicator for whether any household member is covered by health insurance, and an indicator for whether the household has a BPL card. The geographic controls include altitude of the cluster in meters, altitude squared, state level population density (in log) and an indicator whether the place of residence is urban. Health facility controls include hospital per 1000 population and doctors per 1000 population at the state level. Robust standard errors in parentheses clustered at the state level.

## Section C: Robustness to IV Estimates

### Table C1: Alternative Measures of Force Sterilization Policy – Male Sterilization

| | **Panel A: First Stage Estimates**<br>Dependent variable: Excess Male Sterilization (Vasectomy) | | | |
|---|---|---|---|---|
| | (1) | (2) | (3) | (4) |
| Distance from New Delhi to State Capitals (in 100km) | -0.441 | -0.527 | -0.510 | -0.490 |
| | (0.180) | (0.174) | (0.164) | (0.167) |
| Mean of dependent variable | 6.913 | 6.984 | 6.985 | 6.985 |
| F Statistics of Excluded Instrument | 5.98 | 9.18 | 9.63 | 8.62 |
| | **Panel B: Second Stage Estimates**<br>Dependent Variable: Source of Health Care - Public Sector | | | |
| Excess Male Sterilization (Vasectomy) | -0.0369 | -0.0265 | -0.0273 | -0.0285 |
| | (0.0142) | (0.0110) | (0.0110) | (0.00905) |
| Household Controls | NO | YES | YES | YES |
| Geographic Controls | NO | NO | YES | YES |
| Health Facility Controls | NO | NO | NO | YES |
| Observations | 585,634 | 559,899 | 558,755 | 558,755 |
| Mean of dependent variable | 0.448 | 0.443 | 0.442 | 0.442 |

*Notes:* Data are from India's National Family and Health Survey 2015-16 (NFHS-4). The Unit of observation is a household. Household controls include age and sex of the household head, household size, nine religion fixed effects, four caste fixed effects, 21 education of the household head fixed effects, four household wealth index fixed effects, an indicator for whether any household member is covered by health insurance, and an indicator for whether the household has a BPL card. The geographic controls include altitude of the cluster in meters, altitude squared, state level population density (in log) and an indicator whether the place of residence is urban. Health facility controls include hospital per 1000 population and doctors per 1000 population at the state level. Robust standard errors in parentheses clustered at the state level.

Table C2: Alternative Measures of Variation in Forced Sterilization – INC's Vote Share

| | Dependent Variable: Source of Health Care - Public Sector | | | | | | | |
|---|---|---|---|---|---|---|---|---|
| | (1) | (2) | (3) | (4) | (5) | (6) | (7) | (8) |
| INC's Vote Share in 1977 Election | 0.933 | 0.750 | 0.706 | 0.640 | | | | |
| | (0.113) | (0.129) | (0.139) | (0.105) | | | | |
| Change in INC's Vote Share (1977-1971) | | | | | 1.095 | 0.777 | 0.715 | 0.681 |
| | | | | | (0.200) | (0.207) | (0.204) | (0.146) |
| Individual Controls | NO | YES | YES | YES | NO | YES | YES | YES |
| Household Controls | NO | NO | YES | YES | NO | NO | YES | YES |
| Geographic Controls | NO | NO | NO | YES | NO | NO | NO | YES |
| Observations | 456,719 | 437,530 | 437,530 | 437,530 | 349,030 | 332,703 | 332,703 | 332,703 |
| Mean of dependent variable | 0.436 | 0.430 | 0.430 | 0.430 | 0.435 | 0.429 | 0.429 | 0.429 |
| Mean of INC's Vote Share in 1977 Election | 0.384 | 0.382 | 0.382 | 0.382 | | | | |
| Mean of Change in INC's Vote Share (1977-1971) | | | | | -0.158 | -0.160 | -0.160 | -0.160 |

*Notes:* Data are from India's National Family and Health Survey 2015-16 (NFHS-4). The Unit of observation is a household. INC's Vote Share in 1977 Election measures the constituency level variation in INC candidates' vote share in the 1977 election (a proxy measure of excess sterilization). Change in INC's Vote Share (1977-1971) measures the constituency level change in INC candidates' vote share in 1977 compared with the 1971 election (as a second proxy measure of excess sterilization). Household controls include age and sex of the household head, household size, nine religion fixed effects, four caste fixed effects, 21 education of the household head fixed effects, four household wealth index fixed effects, an indicator for whether any household member is covered by health insurance, and an indicator for whether the household has a BPL card. The geographic controls include altitude of the cluster in meters, altitude squared, state level population density (in log) and an indicator whether the place of residence is urban. Health facility controls include hospital per 1000 population and doctors per 1000 population at the state level. Robust standard errors in parentheses clustered at the parliament constituency level.

## Section D: Robustness to Examining the Reasons in NFHS-4

### Table D1: Reasons for Household Generally do not Go to a Government Health Care Facility - Sequential Inclusion of Controls

| | (1) | (2) | (3) | (4) | (5) | (6) |
|---|---|---|---|---|---|---|
| Dependent variable | No nearby facility | Facility timing not convenient | Health personnel often absent | Waiting time too long | Poor quality of care | Other |
| Excess Sterilization | 0.00919 | -0.0274 | -0.00656 | 0.000316 | 0.0614 | 0.0124 |
| | (0.00864) | (0.0127) | (0.00961) | (0.0148) | (0.0151) | (0.00644) |
| Household Controls | NO | NO | NO | NO | NO | NO |
| Geographic Controls | NO | NO | NO | NO | NO | NO |
| Health Facility Controls | NO | NO | NO | NO | NO | NO |
| Observations | 282,333 | 282,333 | 282,333 | 282,333 | 282,333 | 282,333 |
| Mean of dependent variable | 0.446 | 0.264 | 0.148 | 0.409 | 0.479 | 0.0433 |
| Excess Sterilization | -0.000364 | -0.0228 | -0.00555 | 0.00385 | 0.0546 | 0.0114 |
| | (0.00688) | (0.0106) | (0.00863) | (0.00924) | (0.0134) | (0.00532) |
| Household Controls | YES | YES | YES | YES | YES | YES |
| Geographic Controls | NO | NO | NO | NO | NO | NO |
| Health Facility Controls | NO | NO | NO | NO | NO | NO |
| Observations | 274,936 | 274,936 | 274,936 | 274,936 | 274,936 | 274,936 |
| Mean of dependent variable | 0.445 | 0.263 | 0.149 | 0.408 | 0.483 | 0.0440 |
| Excess Sterilization | -0.000959 | -0.0231 | -0.00580 | 0.00364 | 0.0529 | 0.0115 |
| | (0.00626) | (0.00968) | (0.00718) | (0.00865) | (0.0142) | (0.00501) |
| Household Controls | YES | YES | YES | YES | YES | YES |
| Geographic Controls | YES | YES | YES | YES | YES | YES |
| Health Facility Controls | NO | NO | NO | NO | NO | NO |
| Observations | 274,693 | 274,693 | 274,693 | 274,693 | 274,693 | 274,693 |
| Mean of dependent variable | 0.445 | 0.263 | 0.149 | 0.408 | 0.483 | 0.0440 |
| Excess Sterilization | -0.00161 | -0.0193 | -0.00404 | 0.00663 | 0.0596 | 0.00805 |
| | (0.00720) | (0.0103) | (0.00883) | (0.00923) | (0.0163) | (0.00352) |
| Household Controls | YES | YES | YES | YES | YES | YES |
| Geographic Controls | YES | YES | YES | YES | YES | YES |
| Health Facility Controls | YES | YES | YES | YES | YES | YES |
| Observations | 274,693 | 274,693 | 274,693 | 274,693 | 274,693 | 274,693 |
| Mean of dependent variable | 0.445 | 0.263 | 0.149 | 0.408 | 0.483 | 0.0440 |

*Notes:* Data are from India's National Family and Health Survey 2015-16 (NFHS-4). The Unit of observation is a household. Household controls include age and sex of the household head, household size, nine religion fixed effects, four caste fixed effects, 21 education of the household head fixed effects, four household wealth index fixed effects, an indicator for whether any household member is covered by health insurance, and an indicator for whether the household has a BPL card. The geographic controls include altitude of the cluster in meters, altitude squared, state level population density (in log) and an indicator whether the place of residence is urban. Health facility controls include hospital per 1000 population and doctors per 1000 population at the state level. Robust standard errors in parentheses clustered at the state level.

Table D2: Reasons for Household Generally do not Visit a Government Health Care Facility - Male Sterilization

| Dependent variable | No nearby facility | Facility timing not convenient | Health personnel often absent | Waiting time too long | Poor quality of care | Other |
|---|---|---|---|---|---|---|
| | (1) | (2) | (3) | (4) | (5) | (6) |
| Excess Male Sterilization (Vasectomy) | -0.000813 | -0.00973 | -0.00204 | 0.00335 | 0.0301 | 0.00407 |
| | (0.00372) | (0.00576) | (0.00458) | (0.00491) | (0.0112) | (0.00183) |
| Household Controls | YES | YES | YES | YES | YES | YES |
| Geographic Controls | YES | YES | YES | YES | YES | YES |
| Health Facility Controls | YES | YES | YES | YES | YES | YES |
| Observations | 274,693 | 274,693 | 274,693 | 274,693 | 274,693 | 274,693 |
| Mean of dependent variable | 0.445 | 0.263 | 0.149 | 0.408 | 0.483 | 0.0440 |

*Notes:* Data are from India's National Family and Health Survey 2015-16 (NFHS-4). The Unit of observation is a household. Household controls include age and sex of the household head, household size, nine religion fixed effects, four caste fixed effects, 21 education of the household head fixed effects, four household wealth index fixed effects, an indicator for whether any household member is covered by health insurance, and an indicator for whether the household has a BPL card. The geographic controls include altitude of the cluster in meters, altitude squared, state level population density (in log) and an indicator whether the place of residence is urban. Health facility controls include hospital per 1000 population and doctors per 1000 population at the state level. Robust standard errors in parentheses clustered at the state level.

## Section E: Robustness to Confidence in Health Care Facilities and Doctors

### Table E1: Confidence in Institutions: Sequential Inclusion of Controls

| Dependent variable | Confidence: Government hospitals and doctors | | | | Confidence: Private hospitals and doctors | | | |
|---|---|---|---|---|---|---|---|---|
| | (1) | (2) | (3) | (4) | (5) | (6) | (7) | (8) |
| Excess Sterilization | 0.0365 | 0.0379 | 0.0400 | 0.0605 | -0.0376 | -0.0317 | -0.0320 | -0.0325 |
| | (0.0233) | (0.0222) | (0.0233) | (0.0178) | (0.0165) | (0.0143) | (0.0154) | (0.0167) |
| Household Controls | NO | YES | YES | YES | NO | YES | YES | YES |
| Geographic Controls | NO | NO | YES | YES | NO | NO | YES | YES |
| Health Facility Controls | NO | NO | NO | YES | NO | NO | NO | YES |
| Observations | 41,854 | 40,562 | 40,562 | 40,562 | 41,841 | 40,549 | 40,549 | 40,549 |
| Mean of dependent variable | 1.579 | 1.577 | 1.577 | 1.577 | 1.311 | 1.308 | 1.308 | 1.308 |

*Notes:* The data on confidence in hospital and doctors are from IHDS-II. The unit of observation is a household. The IDHS-II assigns the value 1 to "a great deal of confidence," 2 to "only some confidence," and 3 to "hardly any confidence at all." Therefore, higher score constitutes a lower level of confidence. The household controls include household size, income, ten source of main income fixed effects, eight religion fixed effects, five caste fixed effects, two wealth class fixed effects (poor, middle class, (comfortable as reference group)), 16 education of the household head fixed effects, an indicator for whether any household member is covered by government health insurance, an indicator for whether any household member is covered by private health insurance, and an indicator for whether the household has a BPL card. The geographic controls include state level population density (in log) and three place of residence fixed effects. Health facility controls include hospital per 1000 population and doctors per 1000 population at the state level. Robust standard errors in parentheses clustered at the state level.

Table E2: Confidence in Institutions: Alternative Measures of Sterilization

| Dependent variable | Confidence: Government hospitals and doctors | Confidence: Private hospitals and doctors |
|---|---|---|
| | (1) | (2) |
| Excess Male Sterilization (Vasectomy) | 0.0297 | -0.0159 |
| | (0.00846) | (0.00992) |
| Household Controls | YES | YES |
| Geographic Controls | YES | YES |
| Health Facility Controls | YES | YES |
| Observations | 40,562 | 40,549 |
| Mean of dependent variable | 1.577 | 1.308 |

*Notes:* The data on confidence in hospital and doctors are from IHDS-II. The unit of observation is a household. The IDHS-II assigns the value 1 to "a great deal of confidence," 2 to "only some confidence," and 3 to "hardly any confidence at all." Therefore, higher score constitutes a lower level of confidence. The household controls include household size, income, ten source of main income fixed effects, eight religion fixed effects, five caste fixed effects, two wealth class fixed effects (poor, middle class, (comfortable as reference group)), 16 education of the household head fixed effects, an indicator for whether any household member is covered by government health insurance, an indicator for whether any household member is covered by private health insurance, and an indicator for whether the household has a BPL card. The geographic controls include state level population density (in log) and three place of residence fixed effects. Health facility controls include hospital per 1000 population and doctors per 1000 population at the state level. Robust standard errors in parentheses clustered at the state level.